\documentclass[reprint, aps, pre, 10pt, twocolumn, amsmath, amssymb, superscriptaddress]{revtex4-2}
\usepackage{amsfonts}
\usepackage[T1]{fontenc}
\usepackage{multirow}
\usepackage[hidelinks, bookmarks=false]{hyperref}
\usepackage{xcolor}
\hypersetup{
    colorlinks,
    linkcolor={blue!80!black},
    citecolor={blue!80!black},
    urlcolor={blue!80!black}
}

\usepackage{color}

\usepackage{booktabs} 

\usepackage{graphicx}

\usepackage{enumerate}
\usepackage{comment}

\newcommand{\rhobar}{\bar{\rho}}

\newcommand{\hata}{\hat{a}}

\newcommand{\imgi}{\mathrm{\mathbf{i}}}

\newcommand{\lmt}{\tilde{\zeta}}

\usepackage{tikz} 
\newcommand{\fourarrows}{
\begin{tikzpicture}[scale=0.18, baseline={(0,-0.1)}]
    \draw[<->] (-1,0) -- (1,0); 
    \draw[<->] (0,-1) -- (0,1); 
\end{tikzpicture}
}

\newcommand{\be}{\begin{equation}}
\newcommand{\ee}{\end{equation}}
\newcommand{\bea}{\begin{eqnarray}}
\newcommand{\eea}{\end{eqnarray}}

\newcommand{\s}{{\bf s}}

\newcommand{\cQ}{{\cal Q}}

\begin{document}


\title{Generic power laws in higher-dimensional lattice models with multidirectional hopping}

 \author{Animesh Hazra}
 \email{animesh\_edu@bose.res.in}
 \affiliation{Department of Physics of Complex Systems, S. N. Bose National Centre for Basic Sciences, Block-JD, Sector-III, Salt Lake, Kolkata 700106, India.}
 \author{Tanmoy Chakraborty}
 \affiliation{Department of Physics of Complex Systems, S. N. Bose National Centre for Basic Sciences, Block-JD, Sector-III, Salt Lake, Kolkata 700106, India.}
 \author{Anirban Mukherjee}
 \affiliation{Department of Physics of Complex Systems, S. N. Bose National Centre for Basic Sciences, Block-JD, Sector-III, Salt Lake, Kolkata 700106, India.}
 \affiliation{Institute of Physics, Academia Sinica, Taipei 11529, Taiwan.}
 \author{Punyabrata Pradhan}
 \affiliation{Department of Physics of Complex Systems, S. N. Bose National Centre for Basic Sciences, Block-JD, Sector-III, Salt Lake, Kolkata 700106, India.}

\begin{abstract}

We show that, on a $d-$dimensional hypercubic lattice with $d>1$, conserved-mass transport processes, with {\it multidirectional} hopping that respect all symmetries of the lattice, exhibit power-law correlations for generic parameter values $-$ even {\it far} from phase transition point, if any.  
The key idea for generating the algebraic decay is the notion of {\it multidirectional} hopping, which means that several chunks of masses, or several particles, can hop out simultaneously from a lattice site in multiple directions, consequently breaking detailed balance.
Notably, the systems we consider are described by a continuous-time Markov process, are diffusive, {\it lattice-rotation symmetric}, spatially homogeneous and thus have {\it no} net mass current.
Using hydrodynamic and exact microscopic theory, we show that, for spatial dimensions $d > 1$, the steady-state static density-density and ``activity''-density correlation functions in the thermodynamic limit typically decay as $\sim 1/r^{(d+2)}$ at large distance $r=|{\bf r}|$; the strength of the power law is exactly calculated for several models and expressed in terms of the density-dependent bulk-diffusion coefficient and Onsager matrix (or, mobility tensor). 
In particular, our theory explains why center-of-mass-conserving dynamics, used to model novel disordered {\it hyperuniform} state of matter, result in generic long-ranged correlations. However, in a restricted parameter regime, the correlations can also be short ranged and are characterized through the Onsager matrix.

\end{abstract}

\maketitle


\section{Introduction}

Lattice models have played a crucial role in developing a statistical mechanics framework for systems driven out of equilibrium, particularly those having a conserved quantity (e.g., mass) \cite{Spohn1983Dec, Aldous1995Jun, Dickman1998May, Vespignani1998Dec, Majumdar1998Oct}; for reviews, see \cite{Liggett, Bertini2015Jun}.
Yet, most analytical studies rely on approximation techniques \cite{Majumdar1998Oct} and phenomenological field theories \cite{Dickman1998May, Vespignani1998Dec}, and exact results are relatively rare \cite{Dhar1990Apr, Derrida1993Apr, Derrida1998Jan, Derrida2001Sep}.
Indeed, despite decades of intense numerical and theoretical investigations, the precise nature of spatial structures in nonequilibrium systems lacks a good theoretical understanding \cite{Donev2005Aug, Zachary2011Apr, Rissone2021Jul, Kursten2020Feb}.

It is well known that out-of-equilibrium systems exhibit power-law correlations
when translational symmetry is broken $-$ such as in boundary-driven \cite{Spohn1983Dec, Bak1987Jul, Dorfman1994Oct, Dhar2004May, Bertini2015Jun} or disordered systems  \cite{Sadhu2014Jul} $-$ or when rotational symmetry is absent \cite{Grinstein1990Apr, Garrido1990Aug}. 
However, whether such systems can have scale-invariant spatial structures when all underlying (lattice) symmetries are preserved remains an open issue \cite{Bussemaker1996Jun}.
If scale invariance indeed emerge under these symmetries, an important question arises: What classes of nonequilibrium systems are capable of generating such spatial structures? A good theoretical understanding of this issue is still lacking. 
Perhaps not surprisingly, the fact that, in higher dimensions, a wide class of conserved-mass transport processes, e.g., those studied in Refs. \cite{Aldous1995Jun, Dickman2001Oct, Bondyopadhyay2012Jul, Grassberger2016Oct}, possess ``generic scale invariance'' in the presence of all symmetries of the lattice and even far from a phase transition (critical) point is hitherto unknown (presumably due in part to the weak strength of the power-law correlations). 
Indeed, the latter scenario $-$ the appearance of power laws far from the ``critical point'' $-$ does not quite conform to the conventional understanding of critical phenomena, where criticality is achieved at a particular point in the parameter space. This issue is also intriguingly connected to the idea of self-organized criticality (SOC) proposed by Bak, Tang, and Wiesenfeld (BTW) \cite{Bak1987Jul} as an explanation for the scale-invariant structures abundant in nature. 


More specifically, consider, e.g., the Ising model, a paradigm for equilibrium systems with short-ranged interactions, where correlations decay algebraically only at the critical point and short-ranged elsewhere.
Usually, in isotropic systems far from criticality,  the correlation function $C({\bf r})$ for densities at two spatially points separated by position vector ${\bf r} \equiv \{r_{\alpha}\}$ on a $d-$dimensional space, with $\alpha =1, 2, \dots, d$, is short-ranged. Consequently, in the thermodynamic limit, the static structure factor $S({\bf q})$ $-$ Fourier transform of $C({\bf r})$ $-$ as a function of wave-vector ${\bf q} \equiv \{ q_{\alpha} \}$ is analytic and, in the leading order of small $q=|{\bf q}| \to 0$, $S({\bf q}) \simeq S_0(\rhobar) - S_1(\rhobar) |{\bf q}|^2$; here, the two coefficients $S_0(\rhobar)$ and $S_1(\rhobar)$ depend on global density $\rhobar = M/L^d$ with $M$ being total mass (conserved) in the system.
Taking a cue from equilibrium, the prevalent view even in nonequilibrium is that `` ... {\it as for the Ising model, SOC typically exists at a critical point in the relevant parameter space.}'' (see page $1$ in Ref. \cite{Dickman1998May}). That is, closed and isotropic systems 
such as conserved sandpiles \cite{Dickman1998May, Vespignani1998Dec}, are scale-invariant only at the (absorbing) phase transition point, and likely have a finite correlation length otherwise. 
We, however, argue that, unlike in equilibrium, this particular scenario is atypical, and scale invariance is rather generic in nonequilibrium, even in the presence of the lattice rotation and translation symmetries. We characterize a class of lattice models by identifying an interesting feature common to such rotation-symmetric systems $-$ namely, a specific kind of dynamics referred to as {\it multidirectional hopping} $-$ which plays a central role in generating the scale-invariant behavior.

In this paper, we show that higher-dimensional lattice models with mass-conserving and multidirectional hopping have power-law correlations, quite contrary to the expectations, for {\it generic} parameter values and even far from (above) ``critical'' (absorbing-phase transition) point, if any. 
We substantiate our assertion in (i) conserved mass chipping models (MCMs) $-$ generalized versions of random average processes $-$ involving fragmentation, diffusion and aggregation of masses  
\cite{Aldous1995Jun, Krug2000Apr, Rajesh2000May, Bondyopadhyay2012Jul, Chatterjee2014Jan}
and (ii) conserved sandpiles 
\cite{Dickman1998May, Vespignani1998Dec, Rossi2000Aug, Dickman2001Oct, Dhar1990Apr, Basu2012Jul, Hexner2015Mar, Grassberger2016Oct, Wiese2024Aug} $-$ the paradigm for systems undergoing absorbing-phase transition from active to absorbing phase below a critical density. 
These models typically have multidirectional hopping, where {\it multiple} fragments of mass ({\it multiple} particles) at a lattice site hop out simultaneously in {\it several} directions with certain rate and thus violate detailed balance.
They have been extensively studied in the past, but, to our knowledge, have never been reported to have scale invariance far from a phase transition point.

In this paper, using both hydrodynamic and microscopic theories, we show that, on a periodic hypercubic lattice in dimensions $d > 1$, steady-state static density-density and ``activity''-density correlation functions in these systems decay, in the thermodynamic limit and at large distance $r=|{\bf r}|$, as a power law $\sim 1/r^{\eta}$ with exponent $\eta=d+2$; in threshold-activated systems like sandpiles, local ``activity'' at a site indicates whether there can be toppling of ``grains'' or particles (``active'') or not (``inactive'') \cite{Bak1987Jul}. The strength of the power law is calculated exactly for several models and expressed in terms of the bulk-diffusion coefficient and the Onsager matrix (or, the mobility tensor). Strikingly, in models having axial bidirectional hopping of masses, the {\it scaled} ``activity''-density and density-density (two-point) static correlation functions (connected), $C^{{\cal A}m}_{\bf r} = \langle {\cal A}({\bf 0}) m({\bf r}) \rangle_c$ and $C^{mm}_{\bf r} = \langle m({\bf 0}) m({\bf r}) \rangle_c$, respectively, have a generic asymptotic form [see Eq. \eqref{Sq}]; in two dimensions, we have, for ${\bf r} = (x,0)$ with $x$ large,
\begin{align}
\label{eq:scaled_crur_intro}
   \frac{2}{(\gamma_1-\gamma_2)} C^{{\cal A}m}_{x, 0} =\frac{2D}{(\gamma_1-\gamma_2)} C^{mm}_{x, 0} \simeq \frac{6}{\pi x^4},
\end{align}
where $D$ is the bulk-diffusion coefficient and the coefficients $\gamma$'s are related to the Onsager matrix [see Eqs. \eqref{eq:Gqaa} and \eqref{eq:Gqab}, and Fig. \eqref{fig:corrl}].

To demonstrate our arguments, the systems we investigate here are arguably the simplest in the class: They are described by continuous-time Markov processes with nearest-neighbor hopping, and are closed (mass-conserving), {\it diffusive} (reflection-symmetric hopping), spatially homogeneous (translation-invariant) and ``isotropic'' (lattice-rotation-symmetric); of course, they have {\it no} net mass current in the steady state. 
Most crucially, though, they possess multidirectional hopping, which, we demonstrate, leads to {\it nonanalyticity} in static structure factor. 
In other words, under certain symmetry conditions as observed for lattice models on a periodic domain, the emergence of power laws in higher dimensions depend on the violation of detailed balance and the presence of multidirectional hopping.
Indeed, for such systems, we find that the static structure factor for small wave vector ${\bf q} = \{q_{\alpha}\}$ on a $d$-dimensional hypercubic lattice has a generic form,
\be
\label{Sq}
S({\bf q}) \simeq S_0(\rhobar) - S_1(\rhobar) |{\bf q}|^2 + S_2(\rhobar) \frac{\sum_{\alpha=1}^d q_{\alpha}^4}{|{\bf q}|^2},
\ee
where $S_2(\rhobar)$ is another density-dependent coefficient. The structure factor and the correlation functions are ``isotropic'' in the sense that they still have the rotational symmetry of the lattice. However, the higher-order derivatives of the third term in the rhs of Eq. \eqref{Sq} (corresponding to the prefactor $S_2$) are nonanalytic, and results in power-law density correlations $C^{mm}({\bf r}) \sim \int d^d {\bf q} S({\bf q}) e^{-i{\bf q}\cdot{\bf r}} \sim 1/r^{d+2}$ for large $r$.

The rest of the paper is organized as follows. In Sec. \ref{sec:hydro_theoy}, we present a hydrodynamic theory applicable to isotropic diffusive systems in general. In Sec. \ref{sec:models}, we define several nearest-neighbor mass transport models considered in this study. In Sec. \ref{Sec:microgam}, to substantiate the hydrodynamic theory, we provide microscopic calculations for variants of MCMs and threshold-activated systems like sandpiles and density correlation for these models in \ref{sec:correlations}. Finally, in Sec. \ref{sec:conclusion}, we summarize with some concluding remarks.

\section{Hydrodynamic theory}
\label{sec:hydro_theoy}

To physically understand the dynamical origin of the nonanalyticity of the static structure factor, let us simply begin by formulating a linearized hydrodynamic theory of a {\it closed diffusive} system on a periodic domain, albeit out of equilibrium due to violation of detailed balance in (microscopic) configuration space.  
To proceed further, let us define a coarse-grained (fluctuating) density $m(\textbf{r},t)$ at position ${\bf r}$ and time $t$. The density fluctuation $\delta m(\textbf{r}, t) = m(\textbf{r}, t) - \rhobar$ around homogeneous state with global density $\rhobar = \langle m({\bf r}, t \to \infty) \rangle = M/L^d$, satisfy a (discrete) continuity equation due to mass conservation,
\begin{align}
\label{eq:mrt}
    \partial_t \delta m(\textbf{r}, t)  = - \nabla .\left[ {\bf j}^{(d)}(\textbf{r},t) + {\bf j}^{(fl)}(\textbf{r},t) \right],
\end{align}
where $\nabla$ denotes (discrete) gradient and the local instantaneous current ${\bf j} = {\bf j}^{(d)} + {\bf j}^{(fl)}$ in the rhs is decomposed into two parts, with ${\bf j}^{(d)}$ and ${\bf j}^{(fl)}$ being called diffusive and fluctuating (``noise'') currents, respectively. The diffusive current has the following form,
\begin{align}
\label{eq:jdrt}
    {\bf j}^{(d)}(\textbf{r},t) = -\nabla {\cal A}[m(\textbf{r},t)],
\end{align}
where ${\cal A}[m(.)]$ is a coarse-grained fluctuating observable, a generalized ``activity'', which is typically a fast (nonconserved) variable and, in the hydrodynamic framework, can be considered ``slave'' to the local (conserved and slowly varying) density field $m(\textbf{r},t)$ \cite{Tapader2021Mar}; for explicit form of the model-specific quantity $\mathcal{A}$, see microscopic calculations presented in Secs. \ref{sec:Onsager_matrix}. 
Now,  considering small fluctuations $\delta m \ll \rhobar$ and performing a small-gradient expansion, the diffusive current ${\bf j}^{(d)}(\textbf{r},t)$ in Eq. \eqref{eq:jdrt} is written as the gradient (discrete on a lattice) of the density field $m(\textbf{r},t)$, 
\begin{align}
\label{Tr-Sch}
    {\bf j}^{(d)}(\textbf{r},t) \simeq -D(\rhobar) \nabla m(\textbf{r},t),
\end{align}
where we define the bulk-diffusion coefficient $D(\rhobar) = d \langle {\cal A} \rangle/d \rhobar$. In the hydrodynamic framework, the bulk-diffusion coefficient, in principle, depends on local density. But, here it is simply treated as a constant $D(\rhobar)$ with $\rhobar$ being the global density around which the small fluctuations are considered, thus justifying the linearized hydrodynamics discussed below.
The fluctuating current has the properties $\langle j_\alpha^{(fl)}(\textbf{r},t) \rangle =0$, and its strength can be written as
\begin{align}
    \langle j_\alpha^{(fl)}(\textbf{r}’,t’) j_\beta^{(fl)}(\textbf{r}, t)\rangle = \Gamma^{\alpha\beta}(\textbf{r}-\textbf{r}’, \rhobar)\delta(t-t').
\end{align}
Here, the Onsager transport-coefficient matrix $ {\bf \Gamma} \equiv \{ \Gamma^{\alpha\beta}(\textbf{r}-\textbf{r}', \rhobar) \}$ represents the strength of the fluctuating current along any two directions $\alpha$ and $\beta$; the elements of the Onsager matrix are nonzero in general and depend on density and other model-specific parameters.
From microscopic calculations for models considered here, we can analytically calculate the Onsager transport coefficients and thus obtain the strength of the fluctuating (``noise'') current correlation along the same direction $\alpha$,
\begin{align}
\label{eq:Gqaa}
    \Gamma^{\alpha \alpha}_{\textbf{q}}(\rhobar) = \gamma_0(\rhobar) +\gamma_1(\rhobar) \lambda(q_\alpha),
\end{align}
and that along two orthogonal directions $\alpha \ne \beta$,
\begin{align}
\label{eq:Gqab}
    \Gamma_\textbf{q}^{\alpha \beta}(\rhobar) = \gamma_2(\rhobar) (1-e^{\imgi q_\alpha})(1-e^{-\imgi q_\beta}),
\end{align}
where $\lambda(q_\alpha)=2(1-\cos q_\alpha)$. Here we have introduced a set of density-dependent coefficients $\gamma_0$, $\gamma_1$ and $\gamma_2$.
As we show later, for unidirectional mass transfer (or, $1$-particle transfer along a randomly chosen direction), the fluctuating currents along two orthogonal directions are uncorrelated, and we have the following form of $\Gamma^{\alpha\beta} = \gamma_0(\rhobar) \delta_{\alpha,\beta} \delta(\textbf{r}-\textbf{r}')$. However, this particular (delta-correlated) form of  $\Gamma^{\alpha\beta}(\textbf{r}-\textbf{r}')$ is not valid for multidirectional hopping, where cross-correlations between currents along two orthogonal directions also generally appear and moreover $\Gamma^{\alpha \beta}$ has now finite spatial correlations.

We note that, in principle, the hydrodynamic equation \eqref{eq:mrt} involves nonlinear terms, which arise from the density-dependence of the bulk diffusion coefficient $D$ and Onsager matrix $\Gamma^{\alpha \beta}$ $-$ related to the noise strength $-$ thus making the noise multiplicative. 
Indeed, for the specific models studied here, the density-dependence of the relevant transport coefficients appearing in the hydrodynamics are later calculated using a microscopic dynamical approach. Nevertheless, it is still quite instructive to follow the linearized hydrodynamic theory, which, though approximate, however captures the structure factor, and thus the correlation function, remarkably well.

By taking Fourier transform of both sides of  Eq. \eqref{eq:mrt} and 
using Eq.~\eqref{Tr-Sch}, we obtain
\begin{align}\label{eq:demqt}
   \delta m(\textbf{q},t) = \sum_\alpha\int_{0}^{t} dt' e^{-D\omega(\textbf{q})(t-t')}(e^{\imgi q_\alpha}-1)j_\alpha^{(fl)}(\textbf{q},t'),
\end{align}
where $\delta m({\bf q}, t)=\sum_{\bf r}  \delta m(\textbf{r}, t) e^{\imgi \textbf{q} \cdot \textbf{r}}$ and   $\omega(\textbf{q}) = \sum_\alpha \lambda(q_\alpha)$ are the Fourier transforms of excess density $\delta m({\bf r}, t)$ and eigenvalue of the (discrete) Laplacian operator, respectively.
Equation \eqref{eq:demqt} leads to the dynamic structure factor $S(\textbf{q},t) =  L^{-d} \langle | \delta m(\textbf{q},t)|^2\rangle $,
\begin{align}\label{eq:sqt}
    S(\textbf{q},t) =\sum_{\alpha, \beta}\big(1-e^{-2D\omega(\textbf{q}) t}\big)\frac{(1-e^{-\imgi q_\alpha})(1-e^{\imgi q_\beta})\Gamma^{\alpha\beta}_{\textbf{q}}}{2D\omega(\textbf{q})},
\end{align}
and the static structure factor (by taking $t\to\infty$),
\begin{align}
\label{eq:sq}
    S(\textbf{q}) = \frac{1}{2D\omega(\textbf{q})} \sum_{\alpha, \beta} (1 - e^{- \imgi q_\alpha})(1-e^{\imgi q_\beta})\Gamma^{\alpha \beta}_{\textbf{q}} \equiv \frac{ \mathcal{B}(\textbf{q}) }{2D \omega(\textbf{q})},
\end{align}
where we construct a scalar quantity out of the Onsager matrix, 
\begin{align}\label{eq:Bq-Gq}
    \mathcal{B}(\textbf{q}) =  \sum_{\alpha,\beta} (1-e^{-\imgi q_\alpha})(1-e^{\imgi q_\beta})\Gamma^{\alpha\beta}_{\textbf{q}}.
\end{align}
Now, using Eqs. \eqref{eq:Gqaa} and \eqref{eq:Gqab}, we explicitly obtain 
\begin{align}
\label{eq:Bq_all}
    \mathcal{B}(\textbf{q}) = \omega(\textbf{q}) \Bigg[\gamma_0 + \gamma_2 \sum_\alpha \lambda(q_\alpha) + (\gamma_1 -\gamma_2) \frac{\sum_\alpha \lambda^2(q_\alpha)}{\sum_\alpha \lambda(q_\alpha)} \Bigg]
\end{align}
as a function of ${\bf q}$. Indeed, later we resort to a microscopic approach and derive the above expressions for $\mathcal{B}({\bf q})$ and $S({\bf q}, t)$ for all models (MCMs and sandpiles) considered in this paper.
Now, by comparing  Eq. \eqref{Sq} to Eq. \eqref{eq:Bq_all}, one can see that the structure factor can be written in the leading order of small ${\bf q}$ as given in Eq. \eqref{Sq}, and thus we immediately obtain,  
\begin{align}
 &S_0(\rhobar) =\frac{\gamma_0}{2D},
\\
&S_1(\rhobar) = - \frac{\gamma_2}{2D},
\\
&S_2(\rhobar) = \frac{(\gamma_1-\gamma_2)}{2D},\label{eq:s2}
\end{align}
in terms of the transport coefficients $D$ and $\gamma$'s (density-dependent in general).
The above equations \eqref{eq:sq}-\eqref{eq:s2} constitute the main results of this paper.
In fact, the coefficients $D$ and $\gamma$'s can be calculated exactly for MCMs and in terms of the activity for sandpiles (see Table \ref{tab:Gam_bq_table}).

Importantly, we can now see that the third term (i.e., its higher-order derivatives) in Eq. \eqref{Sq}, corresponding to the coefficient $S_2$, is nonanalytic. 
Also, from Eq. \eqref{eq:Bq_all}, we find that the case with $\gamma_1 \neq \gamma_2$ corresponds to nonzero coefficient $S_2 \ne 0$ in Eq. \eqref{Sq} and thus to correlations that exhibit a power-law behavior; otherwise, $S_2=0$ and the correlations are short-ranged.
Furthermore, Eq. \eqref{eq:sq} could be thought of as a modified fluctuation-dissipation relation, connecting the structure factor, density relaxation and current fluctuation in these out-of-equilibrium systems.

\section{Models}
\label{sec:models}

To derive hydrodynamics discussed in the previous section, we consider a broad class of microscopic models, with and {\it without} detailed balance, on a periodic $d-$dimensional hypercubic lattice of volume $L^d$. Notably, the systems have mass-conserving dynamics with {\it symmetric} hopping, which results in diffusive relaxations at large spatio-temporal scales. We consider both unidirectional and multidirectional hopping; also, some of the models we consider have an additional center-of-mass conservation (CoMC), i.e., where both mass and center of mass (CoM) are conserved. The models  evolve through a continuous-time stochastic Markov dynamics as described below.

For simplicity, we define the models (and later present the simulation results) in two dimensions; generalization to higher dimensions is straightforward and performed later in Sec. \ref{sec:correlations}.

\begin{figure}[htbp]
    \centering
    \begin{tikzpicture}[scale=1.2, every node/.style={scale=0.9}]

        \foreach \x in {0,1.2,2.4} {
            \foreach \y in {0,1.2,2.4} {
                \filldraw[black] (\x,\y) circle (1.2pt);
            }
        }

        \draw[->, violet, line width=2pt] (1.2,1.2) -- (1.2,2.4) node at (2.0, 2) {$\lmt m(i, j)\xi^-\xi_y^+$};
        \draw[->, cyan, line width=2pt] (1.2,1.2) -- (1.2,0) node[midway, right] {$\lmt m(i, j)\xi^-\xi_y^-$};
        \draw[->, blue, line width=2pt] (1.2,1.2) -- (2.4,1.2) 
        node at (2.2,1.5) {$\lmt m(i, j)\xi^+\xi_x^+$};
        \draw[->, red, line width=2pt] (1.2,1.2) -- (0,1.2)
        node at (0.2,1.5) {$\lmt m(i, j)\xi^+\xi_x^-$};

        \node at (1,2.8) {(a) MCM I};

        \begin{scope}[xshift=3.8cm]
            \foreach \x in {0,1.2,2.4} {
                \foreach \y in {0,1.2,2.4} {
                    \filldraw[black] (\x,\y) circle (1.2pt);
                }
            }

            \draw[->, blue, line width=2pt] (1.2,1.2) -- (2.4,1.2)
            node at (1.9,1.5) {$\lmt m(i, j) \xi^+$};
            \draw[->, red, line width=2pt] (1.2,1.2) -- (0,1.2)
            node at (0.56,0.9) {$\lmt m(i, j) \xi^-$};
           \draw[->, blue, dotted, line width=2pt] (1.2,1.2) -- (1.2,2.4);
            \draw[->, red, dotted, line width=2pt] (1.2,1.2) -- (1.2,0);
            \node at (1,2.8) {(b) MCM II};
        \end{scope}

    \end{tikzpicture}
    \caption{Schematic representation of dynamical update rules in mass transport processes in two-dimensional space for two variants: (a) MCM I and (b) MCM II. The arrows indicate the direction of mass transfer from the central site $(i,j)$ to its neighbors, with color-coded expressions representing the corresponding fractions of masses transferred during the particular update. Panel (a):  In MCM I, mass is distributed simultaneously to all four nearest neighbors along both $x-$ and $y-$ axes. Panel (b): On the other hand, in MCM II, mass is transferred to only two neighboring sites along either horizontal (solid line) or vertical (dotted line) axis, with equal probability. For other variants (not shown here), when center of mass (CoM) is conserved, the  two chunks of masses transferred from site $(i,j)$, in $+$ve and $-$ve directions along a particular axis, are the same; the respective variants are referred to as MCM-CoMC I and MCM-CoMC II. }
    \label{fig:MCMs}
\end{figure}

\subsection{Unidirectional (one-particle) hopping}


In this class of models,  a {\it single} particle hops from a site $(i,j)$ with rate $u(i, j)$, which depends on the amount of mass at that site. The particle is then attempted to get transferred to one of its nearest neighbors with equal probability ($1/2d$ on a $d$-dimensional square lattice); the move is accepted depending on the presence of hardcore constraint.
For example, in symmetric simple exclusion process (SSEP)  having hardcore exclusion \cite{Derrida2001Sep}, the hop rate is given by $u(i, j) = m(i, j)$, with the occupation variable $m(i, j) \in \{0, 1\}$ taking value $1$ (occupied) or $0$ (vacant), provided the destination site is vacant; otherwise, $u(i, j) = 0$. On the other hand, in the zero-range processes (ZRPs) having unbounded occupation (no hardcore constraint) \cite{Evans2004Jun}, one can have the hop rate, $u(i, j) = 1 + b/m(i, j)$, where mass $m(i, j) \in \{0, 1, 2, \ldots \}$ can take any nonnegative integer value, and $b \ge 0$ is a model parameter. 


\subsection{Multidirectional (multi-particle) hopping}

\subsubsection{Mass chipping models (MCMs)} 

Another class of conserved-mass transport processes  \cite{Bondyopadhyay2012Jul, Das2016Jun, Hazra2024Aug, Hazra2025Feb} we consider here involve fragmentation, diffusion, and coalescence of masses, with total mass in the system being conserved. These models are variants of random average processes \cite{Aldous1995Jun, Ferrari1998Jan, Krug2000Apr}, which belong to the generalized Kipnis-Marchioro-Presutti (KMP) class of models \cite{Kipnis1982Jan, Liu1995Jul, Redig2017Oct, vanGinkel2016Apr, Carinci2013Aug}, and have been intensively studied in the literature over the past decades. 
In these systems, a site $(i, j)$ is updated with unit rate as following. The site retains a constant fraction $\zeta$ of its mass $m(i, j) \geq 0$, while the remaining mass is randomly fragmented and redistributed among its neighbors. 
To proceed further, let us first define a parameter $\lmt=1-\zeta$ (a constant, model parameter) and the random variables $\xi^+ \in (0, 1)$ and $\xi^-=(1-\xi^+)$, where $\xi^+$ is drawn from a uniform distribution (for simplicity), with mean $\mu_1$ and second moment $\mu_2$; similarly, we also define random variables $\xi_x^+$ and $\xi_x^-=(1-\xi_x^+)$, and $\xi_y^+$ and $\xi_y^-=(1-\xi_y^+)$, where $\xi_x^+$ and $\xi_y^+$ are i.i.d and uniformly distributed in unit interval.

In the first variant of MCM, which we refer to as {\it MCM I}, the remaining mass is fragmented into four random parts, $\lmt m(i, j)\xi^+\xi_x^+$, $\lmt m(i, j)\xi^+\xi_x^-$, $\lmt m(i, j)\xi^-\xi_y^+$ and  $\lmt m(i, j)\xi^-\xi_y^-$ and coalesces to four nearest neighbors $(i+1, j)$, $(i-1, j)$, $(i, j+1)$ and $(i, j-1)$, respectively (see Fig. \ref{fig:MCMs}(a)).  
In second variant of MCM, refered to as {\it MCM II}, the chipped-off mass is randomly fragmented into two parts,  $\lmt m(i, j)\xi^+$ and $\lmt m(i, j)\xi^-$,  and then coalesces with either the nearest neighbors in the $x-$direction, $(i+1, j)$ and $(i-1, j)$, or with the nearest neighbors in the $y-$direction,   $(i, j+1)$ and $(i, j-1)$, respectively with equal probability. The mass-transfer rules are schematically presented in Fig. \ref{fig:MCMs}.
Later, we show that both MCM I and II have power-law correlations.

Further, to explore the role of conservation laws, one can think of a variant of MCM I, which, in addition to mass conservation, can now have a center-of-mass conservation (CoMC). In that case, a random fraction of the mass is retained $\xi m(i, j)$, and the remaining fraction is split into four {\it equal} parts (thus imposing CoM conservation on the microscopic dynamics), each of which is then redistributed to each of the four nearest-neighbor sites; we call this CoM-conserving variant as {\it MCM-CoMC I}. In another variant of MCM II with CoMC, simply called {\it MCM-CoMC II}, the chipped-off mass is divided into two equal parts, which are transferred in the opposite directions along either $x-$direction or $y-$direction, chosen with equal probability. Interestingly, while MCM-CoMC I has short-ranged correlations, MCM-CoMC II has power-law correlations.

\subsubsection{Sandpile models }

Here, we consider two variants of conserved (``fixed-energy'') stochastic sandpiles $-$ the Oslo \cite{Christensen1996Jul} and Manna \cite{Manna1991Apr} models, which have been extensively studied in the literature. For simplicity, in the present paper, we study two particular variants of the conserved sandpiles with center-of-mass conservation (CoMC) $-$ the Oslo \cite{Grassberger2016Oct} and Manna \cite{Hexner2017Jan} models with CoMC; 
generalization to other particle transfer rules (say, that without CoMC \cite{Dhar1990Apr, Manna1991Apr} or having continuous mass variables \cite{Basu2012Jul}) is not difficult. In the Oslo model \cite{Grassberger2016Oct}, an active site with mass $m(i, j) \ge m^\star(i,j)$ greater than or equal to a threshold value topples with unit rate: Two particles {\it simultaneously} hop out, with one particle transferred to one its nearest neighbors along a randomly chosen axis, either $(i\pm 1, j)$  or $(i, j\pm 1)$, i.e., axial transfer in opposite directions. Here, the threshold $m^\star(i,j)$ is a random variable that takes value $2$ or $3$; after each toppling, $m^\star(i,j)$ is reset randomly. 
The difference between the Oslo model and the Manna model with CoMC \cite{Hexner2017Jan} is that, in the latter model, $m^\star(i,j)=2$ is fixed, but, in the Oslo model, $m^\star(i,j) \in [2, 3]$ is chosen randomly during a toppling event. 
As discussed later, both the Oslo and Manna models with CoMC have power-law correlations.

Notably, unlike the models studied in Refs. \cite{Maes1990Nov, Maes1991Sep, Garrido1990Aug, Sadhu2014Jul}, hopping rates in the models considered in this work satisfy all symmetries of the lattice. 
However, unlike the equilibrium models such as the SSEP and ZRPs, the MCMs and sandpile models do not possess time-reversal symmetry, meaning they lack detailed balance in the configuration space. Nevertheless, because the hopping rates in the models are ``isotropic'', there is no biasing in any direction and, therefore, the systems have no mass current in the real space. 
The violation of detailed balance  arises because, in MCMs and sandpiles, when multiple particles or chunks of mass simultaneously hop out from a lattice site to several neighboring sites (a forward transition), there is, by definition, no corresponding reverse transition. Therefore the multidirectional hopping implies that the Kolmogorov criterion $-$ a necessary and sufficient condition for detailed balance to hold $-$ is not satisfied in these processes \cite{Kolmogoroff1936Dec, Das2017Jun}. Consequently, unlike in equilibrium systems which have detailed balance, there can now be steady-state probability current in the configuration space, thus driving the systems out of equilibrium.
However, large-scale relaxation dynamics (above absorbing-phase transition point in sandpiles) still remain diffusive. In fact, one can derive and verify an exact diffusion equation (nonlinear in general) that governs the time evolution of the density field in such models \cite{Tapader2021Mar, Hazra2024Aug, Hazra2025Feb}.

\section{Characterization of fluctuations: A Microscopic Approach}
\label{Sec:microgam}


\subsection{Current fluctuations and derivation of hydrodynamics}

In this section, we substantiate the hydrodynamic theory as presented in Sec. \ref{sec:hydro_theoy} by deriving the hydrodynamics from microscopic dynamical calculations.
To this end, we define time-integrated (cumulative) bond current $\cQ_\alpha(\textbf{r}, T)$, along direction $\alpha$, as the net mass flux along the bond between lattice sites $\textbf{r}$ and $\textbf{r}+\hat{e}_\alpha$ over a time interval $T$, where $\hat{e}_\alpha$ is unit vector along $+\alpha$ direction. 
More specifically, consider two mass-transfer events during time interval $T$ where amount of masses $\delta m^{+}$ and $\delta m^-$ are transferred from site ${\bf r}$ to ${\bf r}+\hat{e}_\alpha$ and from site ${\bf r}+\hat{e}_\alpha$ to ${\bf r}$, respectively. Then the cumulative bond current increases by  $\cQ_\alpha(\textbf{r}, T) \rightarrow \cQ_\alpha(\textbf{r}, T) + \delta m^{+} - \delta m^-$.
Accordingly, instantaneous bond current, $j_\alpha(\textbf{r}, t)$, represents the net mass flow per unit time at position $\textbf{r}$ and time $t$.
That is, the time-integrated bond current up to time $t$ along a given direction $\alpha$ can be written as 
\begin{align}
    \cQ_\alpha(\textbf{r}, t) = \int_{0}^{t}dt' j_\alpha(\textbf{r},t'),
\end{align}
where $j_\alpha(\textbf{r},t)$ is
the instantaneous current. Now the local bond current $j_\alpha({\bf r},t)$ can be decomposed into two parts $-$ diffusive (slow or hydrodynamic) and fluctuating (fast or ``noise'') components,
\begin{align}
    j_\alpha({\bf r},t) = j_\alpha^{(d)}(\textbf{r},t) + j_\alpha^{(fl)}(\textbf{r},t),
\end{align}
such that $\langle j_\alpha({\bf r},t) \rangle = \langle  j_\alpha^{(d)}(\textbf{r},t) \rangle$ and therefore $\langle j_{\alpha}^{(fl)} \rangle = 0$.
The first term corresponds to the diffusive current $j_\alpha^{(d)}(\textbf{r},t)$, while the second term represents the fluctuating current $j_\alpha^{(fl)}(\textbf{r},t)$ along direction $\alpha$.

\begin{figure*}
    \centering
    \includegraphics[width=0.29\linewidth]{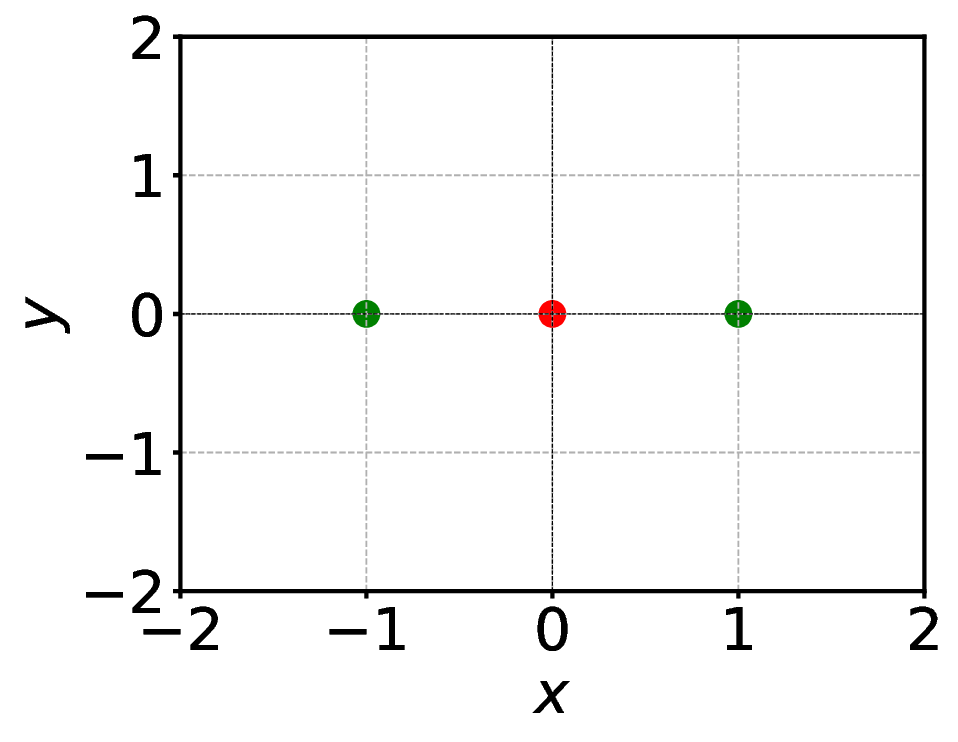}
    \includegraphics[width=0.29\linewidth]{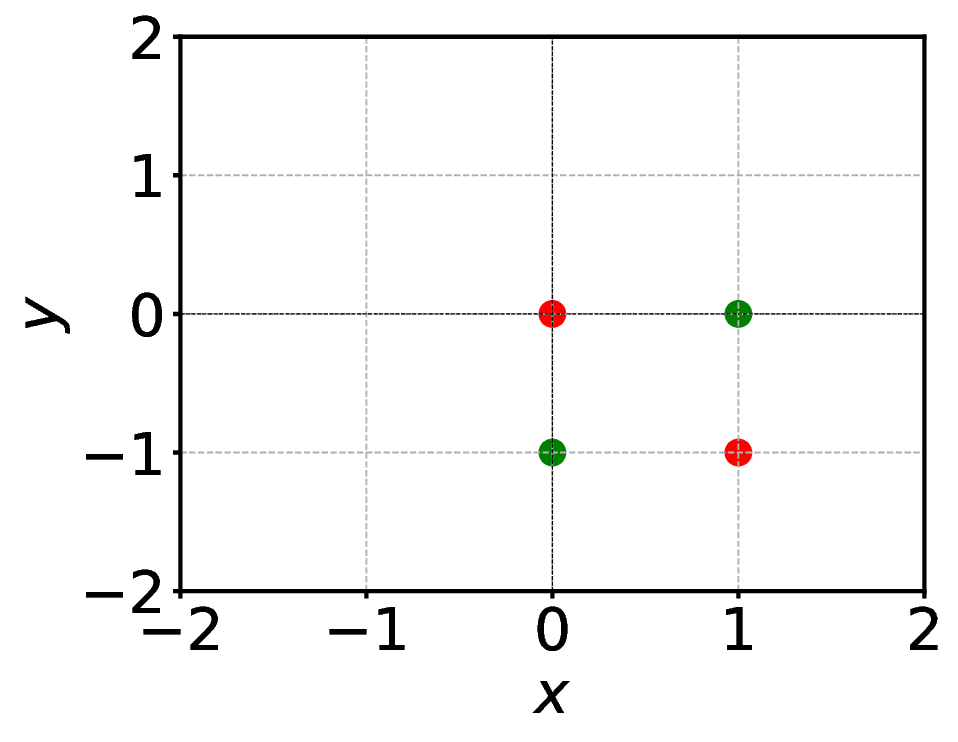}
    \includegraphics[width=0.29\linewidth]{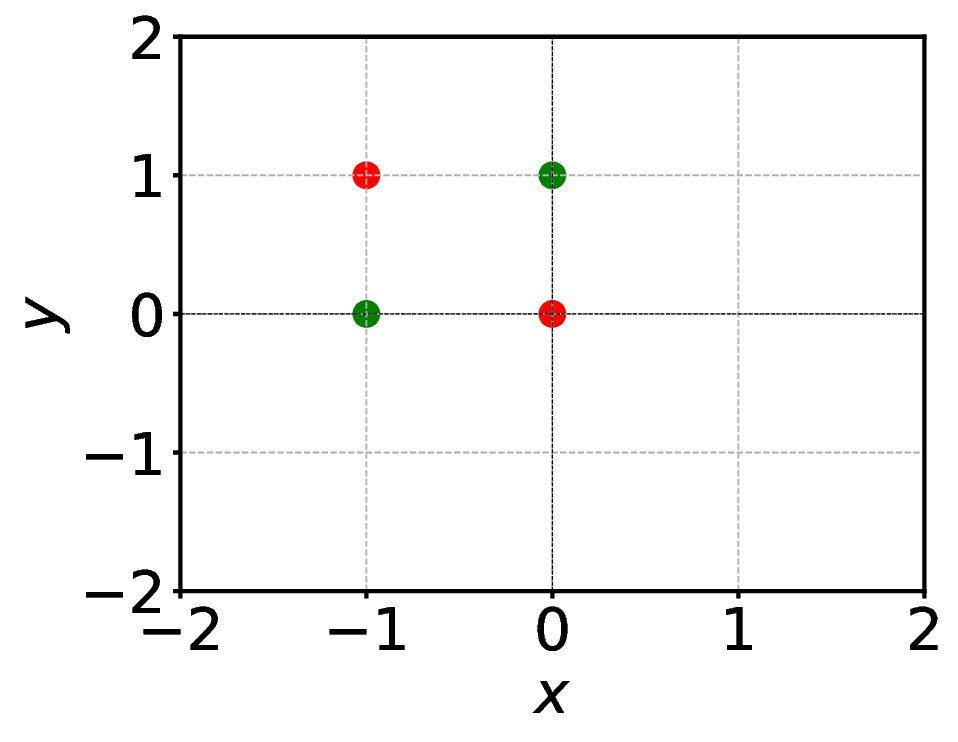}
    \put(-420,100){\textbf{(a)} $\Gamma^{\alpha \alpha}(\textbf{r})$}
    \put(-210,100){\textbf{(b)} $\Gamma^{\alpha \beta}(\textbf{r})$}
    \put(-60,100){\textbf{(c)} $\Gamma^{ \beta\alpha}(\textbf{r})$}
    \caption{\textit{MCM I:} The spatial range of fluctuating current strength is illustrated both along the same direction ($\alpha=x$) (panel (a)) and the perpendicular direction ($\alpha =x$ and $\beta =y$) (panels (b) and (c))  for different nearest-neighbor mass transport processes of the model in $d=2$ considered here. In both panels (a) and (b), the same color represents the same magnitude and sign: red for positive and green for negative values. }
    \label{fig:Gamma_r}
\end{figure*}

For diffusive systems (as considered in this work), the first and second moments of the (time-integrated) bond current are related to the two transport coefficients (density-dependent in general), called the bulk-diffusion coefficient $D$ and the mobility tensor or the Onsager matrix $\Gamma^{\alpha \beta}$ \cite{Bertini2015Jun, Derrida2004May}. These transport coefficients can in general be a tensor quantity, but, for the models considered here, only the latter one is a tensor, while the former one ($D$) is a scalar due to isotropy (lattice-rotation symmetry).  
More specifically, as shown later using microscopic calculations, the average bond current can be written as a gradient of local mass,
\be
\frac{d \langle \cQ_\alpha(\textbf{r}, t) \rangle}{d t} = \langle j_\alpha^{(d)}(\textbf{r},t) \rangle = -\frac{\partial \langle {\cal A} \rangle}{\partial x_{\alpha}} \equiv - D(\rho) \frac{\partial \rho}{\partial x_{\alpha}},
\ee
where we define a density field $\rho({\bf r}, t) = \langle m({\bf r}, t) \rangle$, ${\cal A}({\bf r})$ is a local observable [called ``genaralized activity'' identified later in Eq. \eqref{gen-activity}] and $D(\rho)=d \langle {\cal A} \rangle(\rho)/d \rho$ is the density-dependent bulk-diffusion coefficient. In the last step, we have used a local-equilibrium property where the observable ${\cal A}$ is slave to the density field.
To characterize the second moment, or more generally, the two-point correlations of bond current, we calculate unequal-time $(t>t')$ correlation function for bond current in different directions $\alpha$ and $\beta$ through the following time-evolution equation,
\begin{align}\label{eq:cqcqabttp}
    \frac{d}{dt}C_{\textbf{r}}^{\cQ_\alpha\cQ_\beta}(t,t') = C_{\textbf{r}}^{j^{(d)}_\alpha\cQ_\beta}(t,t'),
\end{align}
where we have defined dynamic correlation as $C_{\textbf{r}}^{\cQ_\alpha\cQ_\beta}(t,t')= \langle \cQ_\alpha(\textbf{0}, t)\cQ_\beta(\textbf{r}, t')\rangle_c$ of two observable $\cQ_\alpha$ and $\cQ_\beta$.
To solve the above equation, we require to obtain equal-time current-current correlation, which satisfies the following equation,
\begin{align}
\label{eq:cqcqabtt}
    \frac{d}{dt}C_\textbf{r}^{\cQ_\alpha\cQ_\beta}(t, t) = C_\textbf{r}^{j^{(d)}_\alpha\cQ_\beta}(t, t)+C_\textbf{r}^{\cQ_\alpha j^{(d)}_\beta}(t, t)+\Gamma^{\alpha\beta}_{\textbf{r}}.
\end{align}
The first two terms in the rhs of the above equation correspond to an infinitesimal change of bond current either in $\alpha$ or in $\beta$ direction, whereas the third term $\Gamma^{\alpha\beta}_{\textbf{r}}$ arise from the simultaneous update of current in both (orthogonal) directions $\alpha$ and $\beta$ [see Eqs. \eqref{eq:Gqaa} and \eqref{eq:Gqab}].  
Now, using some algebraic manipulations, it can be shown that the source term in the above equation is directly related to the strength of the fluctuating current through the Onsager matrix $\Gamma^{\alpha \beta}$ as given below:
\begin{align}
\label{eq:jflabttp}
    C_\textbf{r}^{j^{(fl)}_\alpha j^{(fl)}_\beta}(t,t') = \Gamma^{\alpha\beta}(\textbf{r})\delta(t-t');
\end{align}
see appendix for details. 
In mass transport processes with nearest-neighbor mass transfer (see Table \ref{tab:gamm_all_model}), one can find from the infinitesimal-time update rules for currents that the strength of fluctuating current along the same direction (say, $\alpha$), denoted as $\Gamma^{\alpha \alpha}(\textbf{r})$, is ``positive'' at the origin $\delta(\textbf{r})$ and ``negative'' at the nearest neighbors $\textbf{r} \pm \hat{e}_\alpha$. This leads to the following form in $d$ dimensions,
\begin{align}\label{eq:Gaar_SM}
    \Gamma^{\alpha\alpha}(\textbf{r}) = (\gamma_0+2\gamma_1)\delta(\textbf{r})-\gamma_1[\delta(\textbf{r}+\hat{e}_\alpha)+\delta(\textbf{r}-\hat{e}_\alpha)],
\end{align}
which has the lattice-reflection symmetry $(\textbf{r} \to -\textbf{r})$. Note that, for unidirectional mass transfer (e.g., in the SSEP and ZRPs), we have $\gamma_1 = 0$; on the other hand, in models with center-of-mass conservation, we have $\gamma_0 = 0$.
Furthermore, the strength $\Gamma^{\alpha\beta}(\textbf{r})$ of the fluctuating current along two orthogonal directions $\alpha$ and $\beta$ is interestingly {\it nonzero} if the systems allow multidirectional hopping along orthogonal directions (e.g., MCM I). 
Indeed, we can explicitly write down the analytic expressions for the strength of the cross-correlation functions for fluctuating currents, along two orthogonal directions $\alpha \ne \beta$ in $d$ dimensions as given below:
\begin{align}\label{eq:Gabr_SM}
    &\Gamma^{\alpha\beta}(\textbf{r})= \Gamma^{\beta\alpha}(-\textbf{r})\\ \nonumber &= \gamma_2 [\delta(\textbf{r})-\delta(\textbf{r}-\hat{e}_\alpha)-\delta(\textbf{r}+\hat{e}_\beta)+\delta(\textbf{r}-\hat{e}_\alpha+\hat{e}_\beta)].
\end{align}
Now, after taking Fourier transform of Eqs. \eqref{eq:Gaar_SM} and \eqref{eq:Gabr_SM}, we immediately obtain Eqs. \eqref{eq:Gqaa} and \eqref{eq:Gqab}, respectively.
In Fig. \ref{fig:Gamma_r}, we have schematically represented the spatial range of the fluctuating-current strength along the same direction, $\Gamma^{\alpha\alpha}(\textbf{r})$ 
[ in panel (a)], as well as along the two perpendicular directions,  $\Gamma^{\alpha\beta}(\textbf{r})$  and $\Gamma^{\beta\alpha}(\textbf{r})$ [in panels (b) and (c)].

Furthermore, we define a ``scalar'' (density-dependent) mobility $\chi$ $-$ constructed out of the Onsager matrix $\Gamma^{\alpha \beta}$ $-$ as the variance of the space-time-integrated fluctuating current, integrated over time interval $T$ and over the entire system of volume $V=L^d$. That is, for nearest-neighbor hopping, if $\delta m_{\alpha}$ amount of mass is transferred in $+\alpha$ or $-\alpha$ direction, one adds or subtracts $+ \delta m_{\alpha}$ or $- \delta m_{\alpha}$, respectively, and then considers the variance of the weighted sum of total mass transferred in a time interval $T$. The scaled variance of the weighted sum is related to the mobility $\chi$ through the space-time-integrated currents as 
\begin{widetext}
\begin{align}
\label{eq:2chi_scale}
    2\chi &\equiv \lim \limits_{T\to \infty, V \to \infty} \frac{1}{VTd} \Bigg \langle \left[ \sum_{\alpha, {\bf r}} \cQ_\alpha ({\bf r}, T) \right]^2 \Bigg \rangle   = \lim \limits_{T\to \infty, V \to \infty} \frac{1}{VTd} \Bigg \langle \left[ \sum_{\alpha, {\bf r}} \cQ ^{(fl)}_\alpha ({\bf r},T) \right]^2 \Bigg \rangle 
    \\ \nonumber
    &= \lim \limits_{T\to \infty, V \to \infty} \frac{1}{VTd} \Big \langle \sum_{\alpha, \beta, {\bf r}, {\bf r'}} \cQ ^{(fl)}_\alpha ({\bf r}, T) \cQ^{(fl)}_\beta({\bf r'}, T) \Big \rangle
    = \frac{1}{d}\sum_{\alpha, \beta}\sum_{\textbf{r}}\int_{-\infty}^\infty C_\textbf{r}^{j^{(fl)}_\alpha j^{(fl)}_\beta}(t,0)dt ,
\end{align}
\end{widetext}
where in the first step we used $\cQ_{\alpha} = \cQ_{\alpha}^{(d)} + \cQ_{\alpha}^{(fl)}$ and that the sum over diffusive (gradient) current $\sum_{\bf r} \cQ^{(d)}_{\alpha}({\bf r}, T) \equiv  \int dt \sum_{\bf r} j^{(d)}_{\alpha}({\bf r}, t) = \int dt (-\sum_{\bf r} \partial_{\alpha} {\cal A}) =0$ over a periodic domain vanishes.
Note that the order of limits (of $T$ and $V$) does not matter if one considers integrated current over the entire system. 
Using Eq. \eqref{eq:jflabttp} in the above equation, we can relate the scalar mobility to the spatial correlations involving mass and activity,
\begin{align}
    2\chi = \frac{1}{d}\sum_{\alpha, \beta} \sum_{\textbf{r}} \delta_{\alpha \beta} \Gamma_{\textbf{r}}^{\alpha \beta}=2D\sum_\textbf{r}C_\textbf{r}^{mm}= 2\sum_\textbf{r}C_\textbf{r}^{\mathcal{A}m},
\end{align}
where, in the first step, we have used an identity, for $\alpha\neq \beta,$
\be
\sum_{\textbf{r}} \Gamma_{\textbf{r}}^{\alpha \beta}=0,
\ee
leading to the scalar mobility expressed in terms of the trace of the Onsager matrix,
\begin{align}
\label{eq:2chi_gamma_sm}
    2\chi = \frac{1}{d}\sum_\alpha \sum_{\textbf{r}} \Gamma_{\textbf{r}}^{\alpha \alpha}=\gamma_0.
\end{align}
\begin{figure}
    \centering
    \includegraphics[width=0.8 \linewidth]{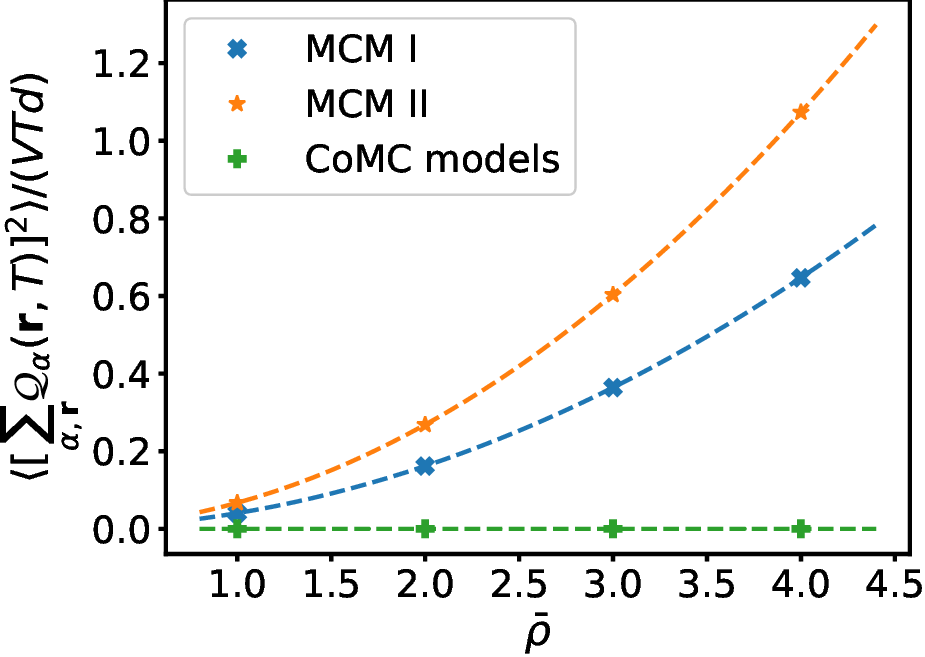}
    \caption{The scaled variance of space-time-integrated currents, $\langle [\sum_{\alpha, \textbf{r}}\cQ_\alpha(\textbf{r}, T)]^2 \rangle/(VTd)$ is plotted as a function of global density $\rhobar$ for MCM I (blue cross), MCM II (orange star), and centre-of-mass conserving models - Manna CoMC and Oslo CoMC (green plus). Points are obtained from simulations, and dotted lines represent $\gamma_0(\rhobar)$ [as in Eq. \eqref{eq:2chi_gamma_sm}; the analytic expressions of $\gamma_0(\rhobar)$ are given in Table \ref{tab:Gam_bq_table} for respective models].}
    \label{fig:tot_fluctuation}
\end{figure}
In Fig. \ref{fig:tot_fluctuation}, we verify the above relation Eq. \eqref{eq:2chi_gamma_sm} by comparing the scaled variance of space-time-integrated current $\langle [\sum_{\alpha, \textbf{r}}\cQ_\alpha(\textbf{r}, T)]^2 \rangle/(VTd) \equiv 2 \chi$ obtained from simulations (points) and the quantity $\gamma_0(\rhobar)$ obtained from theory (dashed lines; see Table \ref{tab:Gam_bq_table}) as a function of global density $\rhobar$ for three different models. 
Both simulations and theory agree with each other quite well.

{\it Fluctuation-dissipation relation.$-$} Interestingly, despite the violation of detailed balance at microscopic scale, there exists a nonequilibrium fluctuation-dissipation (Green-Kubo) relation, analogous to the equilibrium Einstein relation, connecting mass and current fluctuations to density relaxation characterized by the bulk-diffusion coefficient. To derive the relation, we construct a scalar density-dependent transport coefficient, or simply the ``mobility'', out of the Onsager matrix ${\Gamma}^{\alpha \beta}$, 
\begin{align}
    \chi (\rhobar) &\equiv  \frac{1}{2d} \sum_{\alpha, \beta} \delta_{\alpha \beta} \Gamma^{\alpha \beta}_{\textbf{q}\to 0} = \frac{1}{2d} \sum_{\alpha, \beta} \sum_\textbf{r} \Gamma^{\alpha \beta}_{\textbf{r}} 
    \\ \nonumber
    &= D(\rhobar) \sum_\textbf{r}C^{mm}_\textbf{r}(\rhobar) = D S(\textbf{q}\to 0),
\end{align}
where, in the second step, we have used the identity $\Gamma^{\alpha \beta}_{\textbf{q}\to 0}=0$ for $\alpha \ne \beta$ from Eq. \eqref{eq:Gqab}.
In other words, by using Eq. \eqref{Sq}, we have an exact relationship $\chi(\rhobar) = D S_0(\rhobar),$ between the mobility, bulk-diffusion coefficient and scaled mass fluctuation $-$ an Einstein relation for nonequilibrium models studied here.

\subsection{Calculation of the Onsager matrix } \label{sec:Onsager_matrix}

In this section, we calculate the Onsager matrix $\Gamma^{\alpha \beta}$ and express the matrix elements in terms of the coefficients $\gamma$'s as defined in Eq. \eqref{eq:Bq_all}. For simplicity, we consider below only the models in dimensions $d=2$; generalizing the calculations to higher dimensions can be suitably done.

\subsubsection{MCM I in $d=2$ dimensions}

First we exactly calculate the Onsager matrix for a particular variant of MCMs, called MCM I, in $d=2$ dimensions. 
For other variants of MCMs and in higher dimensions, one can proceed along the lines of the calculation scheme provided below.

In MCM I, a random fraction of the chipped-off mass from a site hops to one of its nearest neighbors (for details of the model in dimensions $d=2$, see Sec. \ref{sec:models}).  
We now verify below that the time-evolution equation for the density field indeed has a gradient form \cite{Bertini2015Jun}. To this end, we first write down the time-evolution equation for local mass $m(\textbf{r}, t)$ using the following microscopic infinitesimal-time update rules,
\\\\
$m(\textbf{r}, t+dt)=$
\begin{align}\label{eq:mass_MCMI}
    \begin{cases}
        \textbf{events} & \textbf{prob.} \\
        m(\textbf{r}, t)-\lmt m(\textbf{r}, t) & dt\\
        m(\textbf{r}, t)  + \lmt m(\textbf{r}+\hat{e}_\alpha, t)\xi^+\xi_x^- & dt\\
        m(\textbf{r}, t)  + \lmt m(\textbf{r}-\hat{e}_\alpha, t)\xi^+\xi_x^+ & dt\\
        m(\textbf{r}, t)  + \lmt m(\textbf{r}+\hat{e}_\beta, t)\xi^+\xi_y^- & dt\\
        m(\textbf{r}, t)  + \lmt m(\textbf{r}-\hat{e}_\beta, t)\xi^+\xi_y^+ & dt\\
        m(\textbf{r}, t) & 1- 5dt.
    \end{cases}
\end{align}
From the above update rules, the time-evolution equation for the average local mass can be written as
\begin{align}
    \partial_t \rho({\bf r}, t) = D(\lmt) \nabla^2 \rho({\bf r}, t),
\end{align}
where we denote the density field $\rho({\bf r}, t) = \langle m(\textbf{r},t) \rangle$, $D(\lmt)=\lmt/4$ is the bulk-diffusion coefficient and $\nabla^2$ is the discrete Laplacian operator in $d$ dimensions.
Notably, in this model (and MCMs in general), the bulk-diffusion coefficient $D$ is independent of density, but the mobility (or, equivalently, the coefficients $\gamma$'s) is proportional to the square of the density (see the analytic expressions given in Table \ref{tab:Gam_bq_table}).

\textit{Time evolution of average bond current.$-$} The infinitesimal-time stochastic update rules for bond current $\cQ_{\alpha}(\textbf{r}, t)$ along direction $\alpha \in \{x, y\}$ can be written as\\\\
$\cQ_{\alpha}(\textbf{r}, t+dt)=$
\begin{align}
    \begin{cases}
        \textbf{events} & \textbf{prob.} \\
        \cQ_{\alpha}(\textbf{r}, t)  + \lmt m(\textbf{r}, t)\xi^+\xi_x^+ & dt\\
        \cQ_{\alpha}(\textbf{r}, t)  - \lmt m(\textbf{r}+\hat{e}_\alpha, t)\xi^+\xi_x^- & dt\\
        \cQ_{\alpha}(\textbf{r}, t) & 1- 2dt.
    \end{cases}
\end{align}
Using the above update rule, we get 
\begin{align}
    \partial_t \langle \cQ_{\alpha}(\textbf{r}, t) \rangle= D_\alpha(\lmt) \langle m(\textbf{r}, t) - m(\textbf{r} +\hat{e}_\alpha, t) \rangle \equiv \langle j_{\alpha}^{(d)} \rangle,
\end{align}
from which one can identify the local diffusive current 
\begin{align}
    j_\alpha^{(d)}({\bf r},t) =D_\alpha(\lmt)[ m(\textbf{r}, t) - m(\textbf{r} +\hat{e}_\alpha, t)],
\end{align}
where $D_\alpha(\lmt)= D(\lmt)=\lmt/4$ is the bulk-diffusion coefficient, which is the same (i.e., isotropic) along any direction $\alpha\in \{x, y\}$. 
 Note that, for MCM I (and MCMs in general), we can identify the explicit form of the observable ${\cal A}({\bf r},t) = D(\lmt) m({\bf r},t)$, which enters into the hydrodynamic theory through Eq. \eqref{eq:jdrt}.
Next we characterize the fluctuating current by calculating its strength $\Gamma^{\alpha \beta}$ as in Eq. \eqref{eq:cqcqabtt} or, equivalently, in Eq. \eqref{eq:jflabttp}.

\textit{Time evolution of equal-time bond-current correlations.$-$}
Now we proceed to calculate dynamic bond-current correlations. 
The infinitesimal-time stochastic update rules for equal-time bond current can be written as
\\\\
$\cQ_{\alpha}(\textbf{r}, t+dt) \cQ_\alpha(\textbf{0}, t+dt)=$
\begin{align}
    \begin{cases}
        \textbf{events} & \textbf{prob.} \\
        [\cQ_{\alpha}(\textbf{r}, t)+ \lmt m(\textbf{r}, t) \xi^+ \xi^+_x  ]\cQ_\alpha(\textbf{0}, t)& dt\\
        [\cQ_{\alpha}(\textbf{r}, t)- \lmt m(\textbf{r}+\hat{e}_\alpha, t) \xi^+ \xi^-_x  ]\cQ_\alpha(\textbf{0}, t)& dt\\
        \cQ_\alpha(\textbf{r}, t)[\cQ_{\alpha}(\textbf{0}, t)+ \lmt m(\textbf{0}, t) \xi^- \xi^+_y  ]& dt\\
        \cQ_\alpha(\textbf{r}, t)[\cQ_{\alpha}(\textbf{0}, t)- \lmt m(\hat{e}_\alpha, t) \xi^- \xi^-_y  ]& dt\\
        \cQ_{\alpha}(\textbf{r}, t)\cQ_\alpha(\textbf{0}, t)+ \lmt^2 \{\xi^+\xi^+_x m(\textbf{r}, t)\}^2  &\delta(\textbf{r})dt\\
        \cQ_{\alpha}(\textbf{r}, t)\cQ_\alpha(\textbf{0}, t)+ \lmt^2 \{\xi^+\xi^-_x m(\hat{e}_\alpha, t)\}^2  &\delta(\textbf{r})dt\\
        \cQ_{\alpha}(\textbf{r}, t)\cQ_\alpha(\textbf{0}, t)- (\lmt \xi^+)^2\xi^+_x\xi^-_x m^2(\textbf{r}, t) &\delta(\textbf{r}-\hat{e}_\alpha)dt\\
         \cQ_{\alpha}(\textbf{r}, t)\cQ_\alpha(\textbf{0}, t)- (\lmt \xi^+)^2\xi^+_x\xi^-_x m^2(\textbf{0}, t) &\delta(\textbf{r}+\hat{e}_\alpha)dt\\
        \cQ_{\alpha}(\textbf{r}, t) \cQ_\alpha(\textbf{0}, t) & 1-\Xi dt,
    \end{cases}
\end{align}
where $\Xi=4+2\delta(\textbf{r})+\delta(\textbf{r}+\hat{e}_\alpha)+\delta(\textbf{r}-\hat{e}_\alpha)$ is the total exit rate. Using the above update rules, we obtain the following time-evolution equation for the equal-time bond current correlation function,
\begin{align}\label{eq:cqqtr_MCMI}
    \partial_t C^{\cQ_\alpha\cQ_\alpha}_\textbf{r}(t, t) = 2C^{j^{(d)}_\alpha\cQ_\alpha}_\textbf{r}(t, t)+\Gamma^{\alpha\alpha}(\textbf{r}).
\end{align}
Here the associated source term is given by
\begin{align}\label{eq:Gammar_aa_MCMI}
    \Gamma^{\alpha\alpha}(\textbf{r}) = \frac{\lmt^2 \langle m^2\rangle}{18}[4\delta(\textbf{r})-\delta(\textbf{r}-\hat{e}_\alpha)-\delta(\textbf{r}+\hat{e}_\alpha)],
\end{align}
which can be shown to be equal to the strength of fluctuating current-current correlation [see Eq. \eqref{eq:jflabttp} and the proof given in the Appendix]. Here, the quantity $\langle m^2\rangle$ is the second moment of the onsite mass, which can be calculated by explicitly solving the density correlation function [derived later in Eq. \eqref{eq:cmm_MCMII_sm}].

Now, we write the infinitesimal-time stochastic update rule for bond-current correlations along two orthogonal directions $\alpha$ and $\beta$ as\\\\
\begin{widetext}
\begin{align}
\cQ_{\alpha}(\textbf{0}, t+dt) \cQ_\beta(\textbf{r}, t+dt)=
    \begin{cases}
        \textbf{events} & \textbf{prob.} \\
        [\cQ_{\alpha}(\textbf{0}, t)+ \lmt m(\textbf{0}, t) \xi^+ \xi^+_x  ]\cQ_\beta(\textbf{r}, t)& dt\\
        [\cQ_{\alpha}(\textbf{0}, t)- \lmt m(\hat{e}_\alpha, t) \xi^+ \xi^-_x  ]\cQ_\beta(\textbf{r}, t)& dt\\
        \cQ_\alpha(\textbf{0}, t)[\cQ_{\beta}(\textbf{r}, t)+ \lmt m(\textbf{r}, t) \xi^- \xi^+_y  ]& dt\\
        \cQ_\alpha(\textbf{0}, t)[\cQ_{\beta}(\textbf{r}, t)- \lmt m(\textbf{r}+\hat{e}_\beta, t) \xi^- \xi^-_y  ]& dt\\
        \cQ_{\alpha}(\textbf{0}, t) \cQ_\beta(\textbf{r}, t) + \lmt^2 \xi^+\xi^-\xi^+_x\xi^+_ym^2(\textbf{0}, t)& \delta(\textbf{r})dt\\
        \cQ_{\alpha}(\textbf{0}, t) \cQ_\beta(\textbf{r}, t) - \lmt^2 \xi^+\xi^-\xi^-_x\xi^+_ym^2(\hat{e}_\alpha, t)& \delta(\textbf{r}-\hat{e}_\alpha)dt\\
        \cQ_{\alpha}(\textbf{0}, t) \cQ_\beta(\textbf{r}, t) - \lmt^2 \xi^+\xi^-\xi^+_x\xi^-_ym^2(\textbf{0}, t)& \delta(\textbf{r}+\hat{e}_\beta)dt\\
        \cQ_{\alpha}(\textbf{0}, t) \cQ_\beta(\textbf{r}, t) + \lmt^2 \xi^+\xi^-\xi^-_x\xi^-_ym^2(\hat{e}_\alpha, t)& \delta(\textbf{r}-\hat{e}_\alpha+\hat{e}_\beta)dt\\
        \cQ_{\alpha}(\textbf{0}, t) \cQ_\beta(\textbf{r}, t) & 1-\Xi dt,
    \end{cases}
\end{align}
\end{widetext}
where total exit rate $\Xi=4+\delta(\textbf{r})+\delta(\textbf{r}-\hat{e}_\alpha)+\delta(\textbf{r}+\hat{e}_\beta)+\delta(\textbf{r}-\hat{e}_\alpha+\hat{e}_\beta)$.
Using the above update rules, we immediately obtain the time-evolution equation for time-integrated bond current correlations,
\begin{align}
    \frac{d}{dt}C_\textbf{r}^{\cQ_\alpha\cQ_\beta}(t, t) = C_\textbf{r}^{j^{(d)}_\alpha\cQ_\beta}(t, t)+C_\textbf{r}^{\cQ_\alpha j^{(d)}_\beta}(t, t)+\Gamma^{\alpha\beta}({\textbf{r}}),
\end{align}
where, for $\alpha \ne \beta$, we have
\begin{align}\label{eq:Gammar_ab_MCMI}
    &\Gamma^{\alpha\beta}(\textbf{r}) =\frac{\lmt^2\langle m^2\rangle}{24}
    \\ \nonumber 
    & \times [\delta(\textbf{r})-\delta(\textbf{r}-\hat{e}_\alpha)-\delta(\textbf{r}+\hat{e}_\beta)+\delta(\textbf{r}-\hat{e}_\alpha+\hat{e}_\beta)],
\end{align}
for MCM I in dimensions $d=2$.
 Now, by comparing Eqs. \eqref{eq:Gammar_aa_MCMI} and \eqref{eq:Gammar_ab_MCMI} with Eqs. \eqref{eq:Gaar_SM} and \eqref{eq:Gabr_SM}, we finally obtain the set of coefficients $\gamma$'s as given below:
 \be
 \gamma_0=2\gamma_1=\langle m^2\rangle \frac{\lmt^2}{9},
 \ee 
 and 
 \be 
 \gamma_2= \langle m^2\rangle \frac{\lmt^2}{24}.
 \ee
 As shown later through explicit calculations of density correlation function, we indeed have power-law correlation in MCM I since $\gamma_1 \neq \gamma_2$ (see Table \ref{tab:gamm_all_model} and Fig. \ref{fig:corrl}).
\subsubsection{Oslo Model in $d=2$ dimensions}

In this section, we calculate the Onsager matrix for the two$-$dimensional version of the Oslo model \cite{Grassberger2016Oct}.
The microscopic infinitesimal-time stochastic update rules for local mass can be written as
\\\\
$m(\textbf{r}, t+dt)=$
\begin{align}
\label{eq:mass_Oslo}
    \begin{cases}
        \textbf{events} & \textbf{prob.} \\
        m(\textbf{r}, t)-2 & a(\textbf{r}, t)dt\\
        m(\textbf{r}, t)  + 1& \sum _\alpha a(\textbf{r}+\hat{e}_\alpha, t) {dt}/{2}\\
        m(\textbf{r}, t)  + 1& \sum _\alpha a(\textbf{r}-\hat{e}_\alpha, t) {dt}/{2}\\
        m(\textbf{r}, t) & 1- \Xi dt,
    \end{cases}
\end{align}
where $\Xi=a(\textbf{r}, t)+\sum_{\alpha}\sum_{s\in\{1, -1\}} a(\textbf{r}+s \hat{e}_\alpha, t)/2$  and $a({\bf r}, t)$ is an indicator function if the site ${\bf r}$ is active at time $t$, i.e., $a = 1$ if the site is active, otherwise $a=0$.
Using the above update rules, we straightforwardly obtain the time-evolution equation for the first moment of local mass,
\begin{align}
    \partial_t \langle m(\textbf{r}, t) \rangle = \frac{1}{2}\nabla^2 \langle a(\textbf{r}, t)\rangle .
\end{align}
Here, the rhs of the time-evolution equation is written in terms of the discrete Laplacian of the local average activity. Now, assuming a local-equilibrium (or, local steady state) property where, on large spatio-temporal scale, the local activity is slave to local density $\langle a(\textbf{r}, t) \rangle = \mathsf{a}(\rho({\bf r},t))$ \cite{Bertini2015Jun, Tapader2021Mar}, we obtain the density evolution equation,
\be
    \partial_t \rho(\textbf{r}, t) = \frac{1}{2}\nabla^2 \mathsf{a}(\rho(\textbf{r}, t)),
\ee
where we define density field $\rho({\bf r}, t)=\langle m(\textbf{r}, t) \rangle$ and steady-state activity 
(density of active sites) $\langle \hata({\bf r}) \rangle^{st}_{\rhobar = \rho}  = \mathsf{a}(\rho)$, and $\langle . \rangle^{st}_{\rhobar}$ denotes steady-state average provided that the global density is $\rhobar$.
The local diffusive current $\langle {\bf j}^{(d)}({\bf r}, t) \rangle = - (1/2)\nabla \mathsf{a}(\rho({\bf r}, t)) = - D(\rho) \nabla \rho({\bf r}, t)$ can be immediately identified by casting the above density evolution equation as a continuity equation, where the bulk-diffusion coefficient is given by $D(\rho)=(1/2) d \mathsf{a}/d\rho$, thus justifying diffusive relaxation in Eq. \eqref{Tr-Sch} used in the hydrodynamic theory formulated in the previous section. 


Next, we proceed to write the infinitesimal-time stochastic update rules for equal-time spatial (two-point) correlation function involving time-integrated bond current, \\\\
$\cQ_{\alpha}(\textbf{r}, t+dt) \cQ_\alpha(\textbf{0}, t+dt)=$
\begin{align}\label{eq:qqaa_oslo}
    \begin{cases}
        \textbf{events} & \textbf{prob.} \\
        [\cQ_{\alpha}(\textbf{r}, t)+ 1 ]\cQ_\alpha(\textbf{0}, t)& a(\textbf{r}, t)\frac{dt}{2}\\
        [\cQ_{\alpha}(\textbf{r}, t)- 1  ]\cQ_\alpha(\textbf{0}, t)& a(\textbf{r}+\hat{e}_\alpha, t)\frac{dt}{2}\\
        \cQ_\alpha(\textbf{r}, t)[\cQ_{\alpha}(\textbf{0}, t)+ 1 ]& a(\textbf{0}, t)\frac{dt}{2}\\
        \cQ_\alpha(\textbf{r}, t)[\cQ_{\alpha}(\textbf{0}, t)- 1  ]& a(\hat{e}_\alpha, t)\frac{dt}{2}\\
        \cQ_{\alpha}(\textbf{r}, t)\cQ_\alpha(\textbf{0}, t)+ 1 &a(\textbf{r}, t)\delta(\textbf{r})\frac{dt}{2}\\
        \cQ_{\alpha}(\textbf{r}, t)\cQ_\alpha(\textbf{0}, t)- 1  &a(\textbf{r}, t)\delta(\textbf{r}-\hat{e}_\alpha)\frac{dt}{2}\\
        \cQ_{\alpha}(\textbf{r}, t)\cQ_\alpha(\textbf{0}, t)+ 1 &a(\hat{e}_\alpha, t)\delta(\textbf{r})\frac{dt}{2}\\
         \cQ_{\alpha}(\textbf{r}, t)\cQ_\alpha(\textbf{0}, t)-1 &a(\hat{e}_\alpha, t)\delta(\textbf{r}+\hat{e}_\alpha)\frac{dt}{2}\\
        \cQ_{\alpha}(\textbf{r}, t) \cQ_\alpha(\textbf{0}, t) & 1-\Xi dt,
    \end{cases}
\end{align}
with $\Xi$ is the total exit rate. Using the above update rules, we obtain the following time-evolution equation for dynamic current correlation function, 
\begin{align}\label{dtcqqr_Oslo}
    \partial_t C^{\cQ_\alpha\cQ_\alpha}_\textbf{r}(t, t) = 2C^{j^{(d)}_\alpha\cQ_\alpha}_\textbf{r}(t, t)+\Gamma^{\alpha\alpha}(\textbf{r}),
\end{align}
where we write the strength of the fluctuating current simply as
\begin{align}
    \Gamma^{\alpha\alpha}(\textbf{r}) = \frac{\mathsf{a}(\rhobar)}{2}[2\delta(\textbf{r})-\delta(\textbf{r}+\hat{e}_\alpha)-\delta(\textbf{r}-\hat{e}_\alpha)].
\end{align}
From the above equation, we find that 
\be
\gamma_0  = 0,
\ee 
and 
\be 
\gamma_1 = \frac{\mathsf{a}(\rhobar)}{2}.
\ee 
Notably, the fact that the coefficient \(\gamma_0 = 0\) results in hyperuniform density fluctuation is a consequence of the additional center-of-mass conservation in the Oslo model. Indeed, the dynamical structure of the Manna model with center-of-mass conservation (CoMC) is quite similar to that of the Oslo model and can be characterized similarly (calculations not presented here); see Table \ref{tab:gamm_all_model} for comparison between the two models. 
Now, another coefficient $\gamma_2$ can be calculated from the infinitesimal-time-update rules for bond-current correlations along the orthogonal directions  $\alpha$ and $\beta$ as given below: \\

$\cQ_{\alpha}(\textbf{0}, t+dt) \cQ_\beta(\textbf{r}, t+dt)=$
\begin{align}\label{eq:qqab_oslo}
    \begin{cases}
        \textbf{events} & \textbf{prob.} \\
        [\cQ_{\alpha}(\textbf{0}, t)+ 1 ]\cQ_\beta(\textbf{r}, t)& a(\textbf{0}, t)\frac{dt}{2}\\
        [\cQ_{\alpha}(\textbf{0}, t)- 1  ]\cQ_\beta(\textbf{r}, t)& a(\hat{e}_\alpha, t)\frac{dt}{2}\\
        \cQ_\alpha(\textbf{0}, t)[\cQ_\beta(\textbf{r}, t)+ 1 ]& a(\textbf{r}, t)\frac{dt}{2}\\
        \cQ_\alpha(\textbf{0}, t)[\cQ_\beta(\textbf{r}, t)- 1  ]& a(\textbf{r}+\hat{e}_\beta, t)\frac{dt}{2}\\
        \cQ_{\alpha}(\textbf{0}, t) \cQ_\beta(\textbf{r}, t) & 1-\Xi dt,
    \end{cases}
\end{align}
with $\Xi=[a(\textbf{0}, t)+a(\hat{e}_\alpha, t)+a(\textbf{r}, t)+a(\textbf{r}+\hat{e}_\beta, t)]/2$.
From the above update rules, we obtain time evolution of the equal-time current correlation function,  
\begin{align}
    \frac{d}{dt}C_\textbf{r}^{\cQ_\alpha\cQ_\beta}(t, t) = C_\textbf{r}^{j^{(d)}_\alpha\cQ_\beta}(t, t)+C_\textbf{r}^{\cQ_\alpha j^{(d)}_\beta}(t, t)+\Gamma^{\alpha\beta}({\textbf{r}}),
\end{align}
where, for $\alpha \ne \beta$, we have 
\begin{align}\label{eq:Gammar_ab_oslo}
    \Gamma^{\alpha\beta}(\textbf{r}) =0,
\end{align}
implying $\gamma_2=0$ for the Oslo model. Notably, $\gamma_1 \ne 0$ is still nonzero and, as a result, the system exhibits power-law correlations. However, quite interestingly, the long-wave-length density fluctuations  still vanish in the thermodynamic limit: $S_0=\gamma_0/2D = 0$ as $\gamma_0=0$. That is, unlike systems at an equilibrium critical point, the Oslo model in higher dimensions and even above the (absorbing-phase transition) critical density  exhibits hyperuniformity $-$ suppressed density fluctuations $-$ and long-ranged correlations at the same time.

\subsubsection{Comparison of coefficients \texorpdfstring{$\{ \gamma_0, \gamma_1, \gamma_2 \}$}{gamma} in various models}

As demonstrated in the previous sections, we can also calculate the Onsager matrix $\Gamma^{\alpha\beta}_{\textbf{q}}$ and the quantity $\mathcal{B} (\textbf{q})$ as given  in Eq.~\eqref{eq:Bq_all} for various other mass-transport processes directly from the microscopic dynamics. Indeed, as discussed below, for sandpiles these coefficients can be expressed in terms of the density-dependent activity 
$\mathsf{a}(\rhobar)$ $-$ the ``order parameter'' in the system.
In table \ref{tab:gamm_all_model}, we provide the expressions of the sets of three coefficients $\gamma$'s $\gamma=\{\gamma_0, \gamma_1, \gamma_2\}$, which essentially  encode the information determining whether the mass-mass correlations $C^{mm}_\textbf{r}$, or equivalently activity-mass correlations $C^{\mathcal{A}m}_\textbf{r}$, are power law (PL) or short ranged  (SR).
\begin{table*}[ht!]
\centering
\caption{ {\it Transport coefficients in $d=2$ dimensions.$-$} We provide in this Table the schematic hop directions and the coefficients  $\gamma=\{\gamma_0, \gamma_1, \gamma_2\}$, which parametrize the Onsager transport coefficients $\Gamma^{\alpha \beta}$ (the mobility tensor) for a wide class of conserved-mass transport processes, both with and {\it without} detailed balance, in two dimensions. We have $\langle {\cal A} \rangle \equiv \mathsf{a}(\rhobar)/2$ and $\langle {\cal A} \rangle \equiv u(\rhobar)/2$ $-$ the activity for conserved sandpiles and the average (mass-dependent) hopping rate in the zero-range process (ZRP), respectively; both the quantities depend on the global density $\rhobar$ and other parameters of the models. Also, we denote ``PL'' and ``SR'' as power-law and short-ranged behavior, respectively, of the corresponding correlation functions (for sandpiles, we consider the active phase, i.e., above the critical density where absorbing-phase transition occurs).  }
\label{tab:Gam_bq_table}
\begin{tabular}{l l l l l l}
\toprule
\midrule
\textbf{Models} & Hop direction &$\gamma_0$ & $\gamma_1$ & $\gamma_2$ & Remarks\\
\midrule
I. SSEP / ZRP ($1$-particle hopping) & $[\rightarrow, \leftarrow, \uparrow, \downarrow]$ &$\frac{\rhobar(1-\rhobar)}{2}\big/\frac{u(\rhobar)}{2}$ & $0$ & $0$ & Case 1: SR \\[0.3cm]
II. MCM I ($4$-directional hopping) & $[\fourarrows]$ & $\frac{\lmt^2\pi\rhobar^2}{(9\pi-\lmt(5\pi+2))}$ & $\frac{\lmt^2\pi\rhobar^2}{2(9\pi-\lmt(5\pi+2))}$ & $\frac{3\lmt^2\pi\rhobar^2}{8(9\pi-\lmt(5\pi+2))}$ & Case 3: PL\\[0.3cm]
III. MCM II (axial bidirectional hopping) & $[\leftrightarrow, \updownarrow]$ & $\frac{\lmt^2\rhobar^2\pi}{2(3\pi-\lmt(\pi+4))}$ & $\frac{\lmt^2\rhobar^2\pi}{4(3\pi-\lmt(\pi+4))}$ & $0$ & Case 3: PL \\[0.3cm]
IV. MCM CoMC II (axial bidirectional hopping) & $[\leftrightarrow, \updownarrow]$ & $0$ & $\frac{\mu_1\mu_2\pi\rhobar^2}{8(\pi\mu_1-2\mu_2)}$ & $0$ & Case 4: PL \\[0.3cm]
V. Manna model CoMC \cite{Hexner2017Jan} ($2$-particle hopping) & $[\leftrightarrow, \updownarrow]$ & $0$ & $\frac{\mathsf{a}(\rho)}{2}$ & $0$ &  Case 4: PL\\[0.3cm]
VI. Oslo model \cite{Grassberger2016Oct} ($2$-particle hopping, CoMC) & $[\leftrightarrow, \updownarrow]$& $0$ & $\frac{\mathsf{a}(\rho)}{2}$ & $0$ &  Case 4: PL \\[0.3cm]
\bottomrule
\bottomrule
\end{tabular}
\label{tab:gamm_all_model}
\end{table*}

\subsection{Mass fluctuations: Calculations of two-point spatial correlations}

In this section, we finally calculate the desired two-point correlation functions involving mass (density) and acitivity  and then directly connect the strength of the correlation functions to the Onsager transport coefficients $\Gamma^{\alpha \beta}$ (or, equivalently, $\gamma'$s). To this end, we begin our discussion in a somewhat general setting, where we consider a broad class of microscopic models with multidirectional hopping of masses. For simplicity, here we shall confine ourselves to models with nearest-neighbor hopping of masses (or particles).

During an infinitesimal time interval $(t, t+ dt)$, mass at a site $\textbf{r}$ loses some fraction of the original mass, $\delta m_0(\textbf{r}, t)$, with rate $a(\textbf{r}, t)$ and gains some amount of mass $\delta m_\alpha (\textbf{r}\pm\hat{e}_\alpha)$ with rate $a(\textbf{r}\pm \hat{e}_\alpha, t)$ from a nearest-neighbor site $\textbf{r}\pm \hat{e}_\alpha$ [e.g.,  $\alpha \in (x, y)$ for a $d=2$ dimensional square lattice] with $\hat{e}_\alpha$ is the unit vector along direction $\alpha$. 
Then, the time-evolution equation for density can be written as
\begin{align}\label{eq:mtr_SM}
\partial_t \langle m(\textbf{r}, t)\rangle  &= -\left\langle \delta m_0(\textbf{r}, t)a(\textbf{r}, t) \right\rangle \\\nonumber
&+ \sum_{\alpha=1}^d\sum_{s\in\{1,-1\}} \left\langle \delta m_\alpha(\textbf{r}+s \hat{e}_\alpha, t)a(\textbf{r}+s\hat{e}_\alpha, t)\right\rangle,
\end{align}
where the {\it indicator function} $a(\textbf{r}, t) = \theta_\textbf{r}(m(\textbf{r}, t)-m^\star)$ denotes whether the site ${\bf r}$ is active. For MCMs, $m^*=0$ and $a({\bf r}, t)=1$ at all times. For conserved sandpiles, $m^*$ is called threshold height and takes nonzero integer value(s), depending on the details of the models. That is, if $m({\bf r}) \ge m^*$, $a(\textbf{r}, t)=1$, otherwise it is zero. Provided that a site is updated with unit rate, $a(\textbf{r}, t)$ can essentially be thought of as the (instantaneous) rate with which mass transfer occurs at site $\textbf{r}$ and time $t$. 
The steady-state average of $a(\textbf{r}, t)$, or $ \mathsf{a}(\rhobar)$, depends on global density $\rhobar$, but is independent of ${\bf r}$ for translation-invariant systems considered here.
Also, e.g., for MCM II, the following two quantities $-$ mass loss $\delta m_0(\textbf{r}, t)$ from site ${\bf r}$ and mass gain $\delta m_\alpha(\textbf{r} \pm \hat{e}_\alpha, t)$ from site ${\bf r} \pm \hat{e}_{\alpha}$ $-$ are given by $\lmt m(\textbf{r}, t)$ and $\xi^\mp(\textbf{r} \pm \hat{e}_\alpha, t) \lmt m(\textbf{r} \pm \hat{e}_\alpha, t)$, respectively; in conserved sandpiles with $2$-particle hopping, the corresponding particle (mass) loss and gain are given by $2$ and $1$ for the Manna and Oslo models, respectively.

Now we define a stochastic variable, called generalized ``activity'', as given below:
\be
\label{gen-activity}
{\cal A}(\textbf{r}, t) \equiv \frac{1}{2d} \delta m_0(\textbf{r}, t) a(\textbf{r}, t),
\ee
which is activity $a(\textbf{r}, t)$ weighted by chipped-off mass $\delta m_0(\textbf{r}, t)$.
Therefore, we obtain the time-evolution equation for density field $\rho({\bf r}, t) = \langle m(\textbf{r}, t) \rangle$,
\begin{align}
\label{diff-eqn}
    \partial_t \rho(\textbf{r}, t) = \nabla ^2 \langle{\cal A}(\textbf{r}, t)\rangle = -\nabla \langle \textbf{j}^{(d)}(\textbf{r} ,t) \rangle,
\end{align}
where the local diffusion current can be written in the form of a (discrete)  gradient,
\be 
\label{Fick}
\textbf{j}^{d}(\textbf{r}, t) = -\nabla \mathcal{A}(\textbf{r}, t),
\ee 
a ``gradient''-property \cite{Spohn1983Dec, Bertini2015Jun} satisfied by the models considered in this paper.
In Eq. \eqref{diff-eqn}, we have defined a discrete Laplacian operator as
\begin{align}\label{eq:nabla}
    \nabla ^2 {\cal A}(\textbf{r} ,t) \equiv \sum_{\alpha=1}^d\sum_{s\in\{1,-1\}} {\cal A} (\textbf{r}+s \hat{e}_\alpha, t)-2d {\cal A}(\textbf{r}, t).
\end{align}
Also, as mentioned in Eq. \eqref{eq:jdrt}, Eq. \eqref{Fick} is Fick's law for mass current in the systems.
Note that, on large spatio-temporal scale, we can then define local average diffusive current $\langle\textbf{j}^{(d)}(\textbf{r} ,t)\rangle = - \nabla \langle {\cal A} \rangle (\rho(\textbf{r}, t))$ by invoking a local-equilibrium (local steady state) property where the average $\langle {\cal A} \rangle$ is slave to the density field $\rho({\bf r}, t)$ on large spatio-temporal scale \cite{Bertini2015Jun, Tapader2021Mar}.

Now, using the infinitesimal-time update rules similar to those given explicitly in Eqs. \eqref{eq:mass_MCMI} and \eqref{eq:mass_Oslo},
the time-evolution equation for equal-time density correlation function  $C^{mm}(\textbf{r}_1, \textbf{r}_2, t)=\langle \delta m({\bf r}_1, t) \delta m({\bf r}_2, t)\rangle $, with $\delta m({\bf r}, t) = m({\bf r}, t) - \rhobar$, can be exactly written as
\begin{align}
    \partial_t C^{mm}(\textbf{r}_1, \textbf{r}_2, t)& = \nabla_{\textbf{r}_1} ^2\langle {\cal A}(\textbf{r}_1, t)m(\textbf{r}_2, t)\rangle_c 
    \\ \nonumber
    &+ \nabla^2_{\textbf{r}_2}\langle m(\textbf{r}_1, t) {\cal A}(\textbf{r}_2, t)\rangle_c + B(\textbf{r}_1, \textbf{r}_2).
\end{align}
Since the correlation functions are dependent on relative position vector $\textbf{r}=\textbf{r}_2-\textbf{r}_1$, we have 
\begin{align}
    \partial_t C^{mm}(\textbf{r} ,t)
    = 2\nabla^2 C^{{\cal A}m}(\textbf{r}, t)  + B(\textbf{r}),
\end{align}
where the quantity $B(\textbf{r})$ is a model-specific source term and is strictly short-ranged for models having nearest-neighbor hopping of masses; see Fig \ref{fig:br_range} for a schematic representation of $B({\bf r})$ on a two-dimensional square lattice.

\subsubsection{Calculation of $B(\textbf{r})$ for MCMs in $d=2$ dimensions}

\begin{figure*}[ht!]
    \centering
    \includegraphics[width=1.0\linewidth]{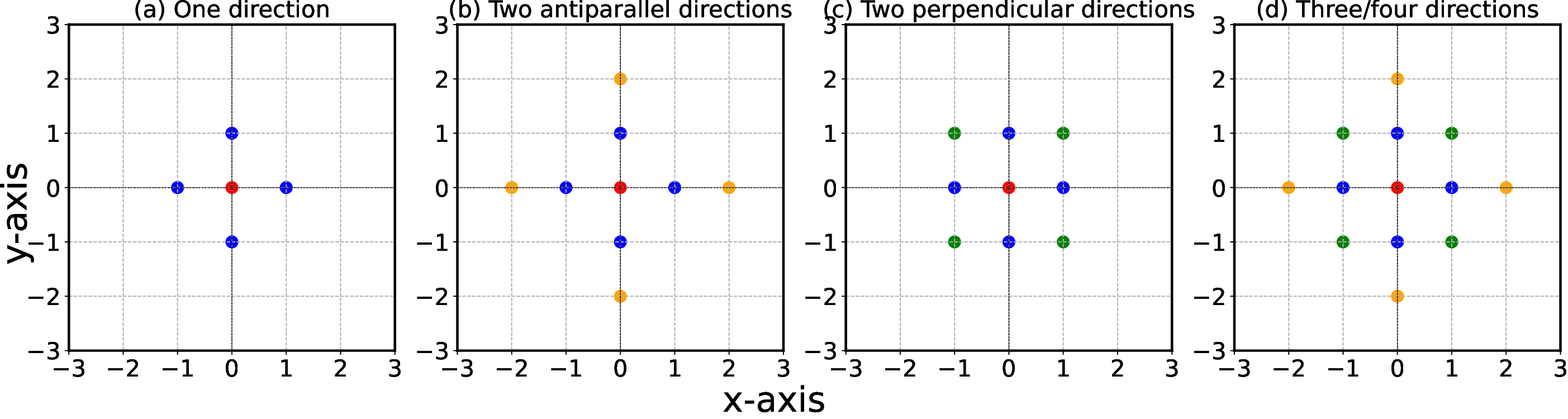}
     \put(-485,120){\textcolor{magenta}{\boldmath$\gamma_1=\gamma_2=0\text{ (SSEP)}$}}
     \put(-340,120){\textcolor{cyan}{\boldmath$\gamma_2=0\text{ (MCM II)}$}}
     \put(-220,120){\textcolor{orange}{\boldmath$\gamma_1=0$}}
     \put(-50,120){\textcolor{teal}{\text{(MCM I)}}}
    \caption{The range of $B(\textbf{r})$ for various nearest-neighbor mass-transfer rules considered in \(d=2\) dimensions in this paper (see Eq. \eqref{eq:Br_rspace}) is analyzed. The quantity \(B(\textbf{r})\) is strictly localized in the sense that it is nonzero on a finite plaque but zero otherwise. Panels (a) (e.g., SSEP, ZRP), (b) (e.g., MCM II, Manna CoMC, Oslo; see Eq. \eqref{eq:BR_oslo}), and (d) (e.g., MCM I; see Eq. \eqref{eq:BR_MCMI}) illustrate the range of \(B(\textbf{r})\) for different models. In panel (c), mass is transported along two randomly chosen perpendicular directions. The same color in each panel represents the same magnitude and sign, as the lattice reflection symmetry \(B(\textbf{r})=B(-\textbf{r})\) holds for the models discussed here.}
    \label{fig:br_range}
\end{figure*}

In this section, we derive the results specifically for MCM I in dimensions $d=2$. Using the infinitesimal-time update rules for local mass in MCM I as given in Eq. \eqref{eq:mass_MCMI}, we obtain the time-evolution equation for the two-point mass-mass spatial correlation function, which satisfies the following equation,
\begin{align}\label{eq:cmtbr_sm}
    \partial_t C^{mm}(\textbf{r}, t) = 2D(\lmt)\nabla^2C^{mm}(\textbf{r}, t)+  B({\bf r}),
\end{align}
where the bulk-diffusion coefficient $D(\lmt)= \lmt/2$ and the source term $B({\bf r})$ can be explicitly written as
\begin{widetext}
\begin{align}\label{eq:BR_MCMI}
    B(\textbf{r}) =  \frac{\lmt^2\langle m^2 \rangle}{36}\left[52\delta(\textbf{r})+\sum_{\substack{\alpha \in \{x, y\} \\ s \in \{-1,1\}}}\{2\delta(\textbf{r}+2s \hat{e}_\alpha)-18\delta(\textbf{r}+s\hat{e}_\alpha)\}+\frac{3}{2}\sum_{\substack{\alpha\neq\beta\\s \in \{-1,1\}}}\{\delta(\textbf{r}+s\hat{e}_\alpha+s\hat{e}_\beta)+\delta(\textbf{r}+s\hat{e}_\alpha-s\hat{e}_\beta)\}\right]
\end{align}
\end{widetext}
with the second moment of mass being denoted as $\langle m^2\rangle \equiv\int m^2\text{Prob.}[m(\textbf{r})=m]dm$. Note that the range of $B(\textbf{r})$ for MCM I is shown in panel (d) of Fig. \ref{fig:br_range}. In steady-state, by putting $\partial_t C^{mm}(\textbf{r}, t)=0$ and then taking the Fourier transform of Eq. \eqref{eq:cmtbr_sm}, we have steady-state structure factor as
\begin{align}\label{eq:sqm2bq_sm}
    S(\textbf{q}) = \frac{\lmt\langle m^2\rangle}{36\omega(\textbf{q})}\left[8 \omega(\textbf{q})+3\omega^2(\textbf{q})+\sum_{\alpha}\lambda^2(q_\alpha)\right],
\end{align}
where $S(\textbf{q})$ is the Fourier transform of steady-state density correction function $C^{mm}(\textbf{r})\equiv C^{mm}(\textbf{r}, t\to \infty)$.
Integrating the steady-state structure factor, as mentioned in Eq. \eqref{eq:sqm2bq_sm}, over the first Brillouin zone, we obtain:
\begin{align}
   C^{mm}(0) \equiv (2\pi)^{-2}\int_{BZ}S(\textbf{q})d\textbf{q} =  (C^{mm}(0)+\rhobar^2)
   \nonumber \\
   \times (2\pi)^{-2}\frac{\lmt}{36}\int_{BZ} d\textbf{q}\left[8+3\omega(q)+\frac{\sum_{\alpha\in \{x, y\}}\lambda^2(q_\alpha)}{\omega(\textbf{q})}\right]
\end{align}
where we have used $\langle m^2 \rangle = (C^{mm}(0)+\rhobar^2)$. After performing some algebraic manipulations, we do the above integral and obtain the variance of onsite mass as
\begin{align}\label{eq:cmm_MCMII_sm}
    \langle m^2\rangle= \frac{9\pi\rhobar^2}{9\pi-\lmt(5\pi+2)}.
\end{align}
Using above equation in \eqref{eq:sqm2bq_sm}, we have
\begin{align}\label{eq:Bq_MCMI}
    \mathcal{B}(\textbf{q}) =  \frac{\lmt^2\pi\rhobar^2}{8(9\pi-\lmt(5\pi+2))} \left[8\omega(\textbf{q}) +3\omega^2(\textbf{q}) +\sum_\alpha\lambda^2(q_\alpha) \right].
\end{align}
The above expression is a special case of Eq. \eqref{eq:Bq_all}, where the model-dependent coefficients $\gamma$'s are nonzero. It is worth noting that $\mathcal{B}(\textbf{q})$ can also be obtained from the fluctuating relation
\begin{align}
    \mathcal{B}(\textbf{q}) = \sum_{\alpha, \beta} (1 - e^{- \imgi q_\alpha})(1 - e^{\imgi q_\beta})\Gamma^{\alpha \beta}_{\textbf{q}},
\end{align}
by knowing the fluctuating current strength $\Gamma^{\alpha \beta}_{\textbf{q}}$ as mentioned in Eq. \eqref{eq:sq}.
In other words, in the real space, the quantities $B({\bf r})$ and $\Gamma^{\alpha \beta}({\bf r})$ are related through  derivatives (discrete on a lattice)
\begin{align}
\label{eq:brgamr}
    B(\textbf{r}) = -\sum_{\alpha, \beta}\partial_\alpha\partial_\beta\Gamma^{\alpha\beta}(\textbf{r}),
\end{align}
can be thought of as a fluctuation relation connecting static and dynamic fluctuations, i.e., mass and current correlations, in the system.
Indeed, from the microscopic infinitesimal-time update rules, we can explicitly show that the above relation holds for other models considered here and is quite generic for diffusive systems with nearest-neighbor mass transfer; see Table \ref{tab:gamm_all_model}.
Equivalently, using Eqs. \eqref{eq:Gaar_SM} and \eqref{eq:Gabr_SM}, one can write down the following general expression for $B(\textbf{r})$ in $d=2$ dimensions,
\begin{widetext}
\begin{align}\label{eq:Br_rspace}
    B(\textbf{r}) =  (4\gamma_0+12\gamma_1+8\gamma_2)\delta(\textbf{r})&+ \sum_{\substack{\alpha\in\{x, y\}\\s \in \{-1,1\}}}\{-(\gamma_0+4\gamma_1+4\gamma_2)\delta(\textbf{r}+s\hat{e}_{\alpha})+\gamma_1\delta(\textbf{r}+2s\hat{e}_\alpha)\}\nonumber \\  
    &+\gamma_2\sum_{\substack{\alpha\neq\beta\\s \in \{-1,1\}}}\{\delta(\textbf{r}+s\hat{e}_\alpha+s\hat{e}_\beta)+\delta(\textbf{r}+s\hat{e}_\alpha-s\hat{e}_\beta)\},
\end{align}
\end{widetext}
explicitly in terms of the coefficients $\gamma$'s, which depend on global density and other parameters of the models. The Fourier transform of the above equation is nothing but the expression of $B({\bf r})$ as given in Eq. \eqref{Sq}. In Fig. \ref{fig:br_range}, we plot the spatial range of  $B(\textbf{r})$ for different values of $\gamma$'s corresponding to various models in $d=2$ dimensions (see Table \ref{tab:gamm_all_model}).

\subsubsection{Calculation of $B({\bf r})$ in sandpiles in $d=2$ dimensions}

Let us first consider the conserved Oslo model in $d=2$ dimensions.
Using the infinitesimal-time stochastic update rules for local mass as given in Eq. \eqref{eq:mass_Oslo}, the time-evolution equation for two-point mass-mass correlation function can be written as
\begin{align}
    \partial_t C^{mm}(\textbf{r}, t) = 2\nabla^2 C^{\mathcal{A}m}(\textbf{r}, t) + B(\textbf{r}),
\end{align}
where the source term is given by
\begin{align}\label{eq:BR_oslo}
    &B(\textbf{r})\nonumber\\& = \frac{\mathsf{a}(\rhobar)}{2}\Big[12\delta(\textbf{r})+\sum_{\substack{\alpha\in\{x, y\}\\s \in \{-1,1\}}}\{\delta(\textbf{r}+2s\hat{e}_\alpha)-4\delta(\textbf{r}+s\hat{e}_\alpha)\}\Big].
\end{align}
Here, the quantity $\mathsf{a}(\rhobar)$ is the steady-state activity, which is a function of global density $\rhobar$. 
Notably, for the Manna model with center-of-mass conservation \cite{Hexner2017Jan}, the equation for $B({\bf r})$ has exactly the same structure, with $\mathsf{a}(\rhobar)$ being the density-dependent activity in the respective system. Also, it is not difficult to show that the relationship between $B({\bf r})$ and $\Gamma^{\alpha \beta}({\bf r})$ remains the same as given in Eq. \eqref{eq:brgamr}. 


\subsubsection{Correlation functions in $d$ dimensions}\label{sec:correlations}

In this section, we derive the correlation functions in arbitrary spatial dimensions $d > 1$, find the large-distance asymptotics and then verify the results by   in dimensions $d=2$ through simulations.

Irrespective of the specific details of the models considered in this work, we can write down the equations for static spatial correlation function involving  density (mass) and local activity. 
By using Eqs. \eqref{eq:jdrt} and \eqref{Tr-Sch} and then taking inverse Fourier transform of Eq. \eqref{eq:sq}, we find that the static two-point density-density and activity-density correlation functions, $C^{mm}(\textbf{r}) = \langle \delta m({\bf 0})\delta m({\bf r}) \rangle$ and $C^{{\cal A} m}(\textbf{r}) = \langle \delta {\cal A}({\bf 0}) \delta m({\bf r})\rangle$, respectively, in the real space satisfy a Poisson equation, 
\begin{align}
\label{eq:C-r}
    2 \nabla^2 C^{{\cal A} m}(\textbf{r}) +  B(\textbf{r}) = 0,
    \\ \label{eq:C-rb}
    2 D \nabla^2 C^{mm}(\textbf{r}) + B(\textbf{r}) = 0,
\end{align}
where $B(\textbf{r})$, the inverse Fourier transform of $\mathcal{B}(\textbf{q})$, can be thought of as a charge distribution in an electrostatic problem. Also, from mass conservation and rotation symmetry, we have $\int B({\bf r}) d {\bf r}=0$ and $\int {\bf r} B({\bf r}) d {\bf r}=0$ (analogous to quadrupole charge distribution). 
Note that the Eq. \eqref{eq:C-r} involving activity-density correlation is exact for both MCMs and sandpiles (in {\it active} phase). The second equation \eqref{eq:C-rb} is exact for MCMs.
For sandpiles, to arrive at Eq. \eqref{eq:C-rb}, we have used an approximation $\nabla C^{{\cal A} m}(\textbf{r}) = \langle \nabla \delta {\cal A}({\bf r}) \delta m({\bf 0})\rangle \simeq D(\rhobar) \langle \nabla \delta m({\bf r}) \delta m({\bf 0})\rangle = D(\rhobar)  \nabla C^{mm}(\textbf{r})$. 
Here, we have essentially assumed that, on large spatio-temporal (hydrodynamic) scales, activity ${\cal A}({\bf r}) = {\cal A}[m({\bf r})]$ is slave to the local density field $m({\bf r})$ \cite{Tapader2021Mar}.

We remark here that, for MCMs, both equations are exact and essentially encode the same information.
Also, in all models considered here, the fluctuating-current correlation function $\Gamma^{\alpha \beta}(\textbf{r})$ 
has been calculated exactly from the infinitesimal-time microscopic dynamics of currents and masses at two different space points with spatial separation $r=|{\bf r}|$. As shown through the explicit microscopic  calculations in the previous sections, the source term $B(\textbf{r})$ appearing in Eqs. \eqref{eq:C-r} and \eqref{eq:C-rb} is indeed directly related to the Onsager transport coefficients $\Gamma^{\alpha \beta}(\textbf{r})$ through a fluctuation-response relation given in Eq. \eqref{eq:brgamr}. 
It is worth mentioning that, although the quadrupole moments (diagonal part) of $B({\bf r})$ 
are nonzero in these models, they are all equal due to the isotropic nature of the systems. Consequently, 
the large-distance $1/r^{(d+2)}$ power-law behavior we observe are generated by the higher-order (octupole) moments of $B({\bf r})$, not the quadrupole ones, which are known to govern the $1/r^d$ algebraic decay of the two-point density correlation function in anisotropic models \cite{Garrido1990Aug}.

Now, taking Fourier transform of Eqs. \eqref{eq:C-r} and \eqref{eq:C-rb}, we explicitly obtain the density-density and activity-density correlation function, in terms of density and all other parameters, 
\begin{subequations}\label{eq:LRC}
\begin{align}
    &C^{{\cal A} m}(\textbf{r}) =  \frac{1}{(2\pi)^d} \int_{BZ} d^d\textbf{q} e^{-\imgi \mathbf{q} \cdot \mathbf{r}} \frac{\mathcal{B}(\textbf{q})}{2 \omega(\textbf{q})},\\
    &C^{mm}(\textbf{r}) 
    =  \frac{1}{(2\pi)^d} \int_{BZ} d^d\textbf{q} e^{-\imgi \mathbf{q} \cdot \mathbf{r}} \frac{\mathcal{B} (\textbf{q})}{2D\omega(\textbf{q})},
\end{align}
\end{subequations} 
where we have essentially performed the inverse Fourier transform of the structure factor $S({\bf q})$, by taking the thermodynamic limit ($L \to \infty$) and performing the ${\bf q}$-integration over the first Brillouin zone (BZ). 
Using the general expression of $\mathcal{B}(\mathbf{q})$ from Eq.\eqref{eq:Bq_all} in Eq.\eqref{eq:LRC}, we obtain the correlation function in real space,
\begin{align}
    C^{\mathcal{A}m}(\textbf{r}) &=DC^{mm}(\textbf{r})= \frac{\gamma_0+2d\gamma_2}{2}\delta(\textbf{r})\\ \nonumber &-\gamma_2\sum_{\alpha=1}^d\sum_{s\in [-1,1]}\delta(\textbf{r}+s\hat{e}_\alpha)+\frac{\gamma_1-\gamma_2}{2}C_{LR}(\textbf{r}).
\end{align}
Here, the first two terms in the rhs of the above equation refer to the short-ranged spatial correlations; the last term contains the power-law decay, which can be expressed as,
\begin{align}\label{eq:clr_app}
 C_{LR}(\textbf{r})&=\frac{1}{(2\pi)^{d}}\int_{BZ}d\textbf{q}\frac{\sum_\alpha\lambda^2(q_\alpha)}{\sum_\alpha\lambda(q_\alpha)}e^{-\imgi \mathbf{q} \cdot \mathbf{r}}
\end{align}
where $\lambda(q_\alpha)= 2(1-\cos q_\alpha)$ is the eigenvalue of the Laplacian operator in the lattice along $\alpha$ direction, and the Brillouin zone is $[-\pi, \pi]^d$.
We now express the numerator $\sum_\alpha \lambda^2(q_\alpha)$ as the Fourier transform of a real-space kernel: 
\begin{align}\label{eq:Kr}
    \sum_\alpha \lambda^2(q_\alpha)= \sum_{\textbf{r}'} K(\textbf{r}')e^{\imgi \textbf{q}\cdot \textbf{r}'},
\end{align}
where $K(\textbf{r}')$ can be explicitly written as
\begin{align}\label{eq:KR1}
\nonumber K(\textbf{r}')&=6d\delta(\textbf{r}')-4\sum_{\alpha=1}^d[\delta(\textbf{r}'+\hat{e}_\alpha)+\delta(\textbf{r}'-\hat{e}_\alpha)]\\ 
&+\sum_{\alpha=1}^d[\delta(\textbf{r}'+2\hat{e}_\alpha)+\delta(\textbf{r}'-2\hat{e}_\alpha)].
\end{align}
Using \eqref{eq:Kr} in \eqref{eq:clr_app}, we have 
\begin{align}
\label{eq:CLR1}
    C_{LR}(\textbf{r}) = \sum_{\textbf{r}'} K(\textbf{r}')\mathcal{G}(\textbf{r}-\textbf{r}') \simeq \sum_{\alpha=1}^d \frac{\partial^4}{\partial x_\alpha^4}\mathcal{G}(\textbf{r}),
\end{align}
where $\mathcal{G}(\textbf{r})$ is the lattice Green's function,
\begin{align}
    \mathcal{G}(\textbf{r}) = \int_{BZ}\frac{d^d\textbf{q}}{(2\pi)^d} \frac{1}{\sum_\alpha \lambda(q_\alpha)} e^{-\imgi \textbf{q}\cdot \textbf{r}}.
\end{align}
After some algebraic manipulation, the above equation can be expressed in the following form,
\begin{align}\label{eq:GR_1}
    \mathcal{G}(\textbf{r}) = \int_0^\infty dt e^{-2td}\Pi_{\alpha}\mathcal{I}_{x_\alpha}(2t),
\end{align}
where we have used the identity $1/A = \int_0^\infty e^{-A t}dt$ with $A=\sum_\alpha \lambda(q_\alpha)$ and have defined
\begin{align}
\label{eq:bessel}
\mathcal{I}_{x_\alpha}(2t)  = \frac{1}{2\pi} \int_{-\pi}^{\pi} dq_\alpha e^{2t \cos q_\alpha}  e^{\imgi q_\alpha x_\alpha},
\end{align}
the modified Bessel function of the first kind \cite{abramowitz1965handbook}.\\\\
\textit{Large-Distance Behavior.$-$} For distance large, the asymptotic behavior of the above integral in Eq.\eqref{eq:bessel} is governed by the small $\textbf{q}$ behavior of the integrand and the corresponding asymptotics can be evaluated as follows. By using the approximation $\cos q_\alpha \simeq (1-q_\alpha^2/2)$ for small $q_{\alpha}$, Eq. \eqref{eq:bessel} has following asymptotic form for large $x_{\alpha} \gg 1$,
\begin{align}\label{eq:bessel_asym1}
    \mathcal{I}_{x_\alpha}(2t) \simeq\frac{e^{2t}}{2\pi}\int_{-\infty}^\infty dq_\alpha e^{-tq_\alpha^2}e^{-\imgi q_\alpha x_\alpha}=\frac{e^{2t}}{\sqrt{4\pi t}}e^{-x_\alpha^2/4t},
\end{align}
which can then be substituted 
in Eq. \eqref{eq:GR_1} to obtain the following equation,
\begin{align}
    \mathcal{G}(\mathbf{r}) \simeq \frac{1}{(4\pi)^{{d}/{2}}}\int_0^{\infty}dt e^{-{r^2}/{4t}} t^{-{d}/{2}}.
\end{align}
By substituting the above equation in Eq.\eqref{eq:CLR1}, we find that, in higher  dimensions $d \ge 2$, the long-range part of the correlation function can be expressed as:
\begin{align}
\nonumber C_{LR}(\textbf{r})&\simeq \sum_\alpha \frac{\partial^4}{\partial x_\alpha^4} \mathcal{G}(\mathbf{r})
\\ 
&\simeq \frac{1}{(4\pi)^{d/2}} \int_0^\infty dt e^{-{r^2}/{4t}} \frac{(\sum_\alpha x_\alpha^2+12dt^2-12 r^2t)}{16t^{4+{d}/{2}}} 
\\
&= \frac{1}{\Omega_d}\frac{d(d+2)(d+4)\sum_\alpha x_\alpha^4-3d(d+4)r^4}{r^{d+6}},
\end{align}
where $\Omega_d=2\pi^{d/2}/\Gamma(d/2)$ is the solid angle in $d$ dimensions. For $r \gg 1$ being large, we have $C_{LR}(\textbf{r}) \sim 1/r^{d+2}$; the algebraic decay is valid for $d>1$, as seen in the above equation. 
Putting $d=2$ and performing some algebraic manipulations, the above equation leads to the following expression of the long-ranged correlation in two dimensions,
\begin{align}
    C_{LR}(x, y) = \frac{6 \left(x^4 - 6 x^2 y^2 + y^4\right)}{\pi \left(x^2 + y^2\right)^4}.
\end{align}
Thus, in the case of a two-dimensional system, the scaled correlation functions are given by
\begin{align}
\label{eq:scaled_crur}
   \frac{2}{(\gamma_1-\gamma_2)}C^{{\cal A} m}_{x, 0}=\frac{2D}{(\gamma_1-\gamma_2)}C^{mm}_{x, 0} \simeq \frac{6}{\pi x^4},
\end{align}
in the limit of distance ($1 \ll |{\bf r}| \ll L$) being large. However, in a finite system,  the power law is essentially cut-off at very large $|{\bf r}|$ with $|{\bf r}|/L$ being finite, and there are finite-size corrections of ${\cal O}(1/L^d)$ to the correlation functions.

\begin{figure}[!ht]
    \centering
    \includegraphics[width=0.9\linewidth]{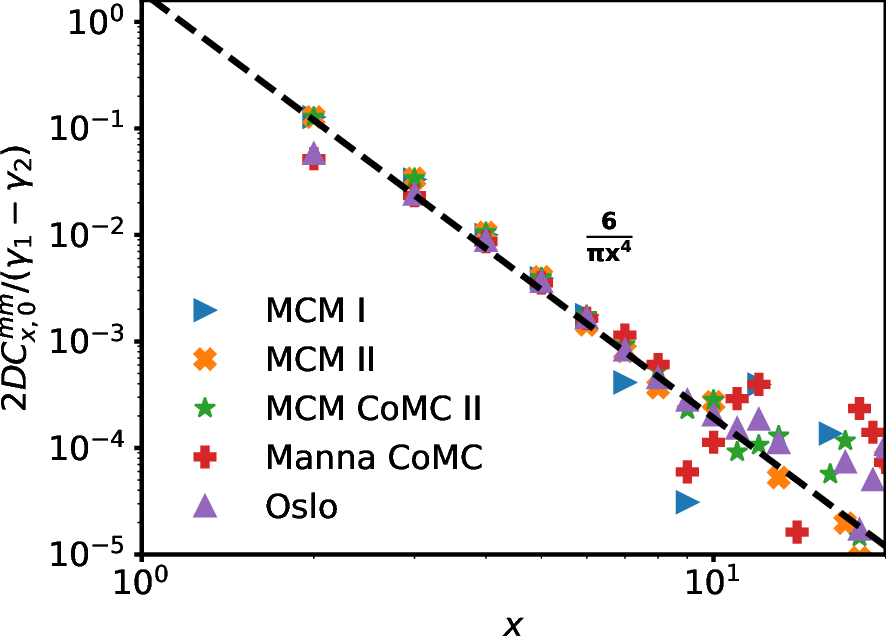}
    \put(-100,150){\textbf{(a)}}\\
    \includegraphics[width=0.9\linewidth]{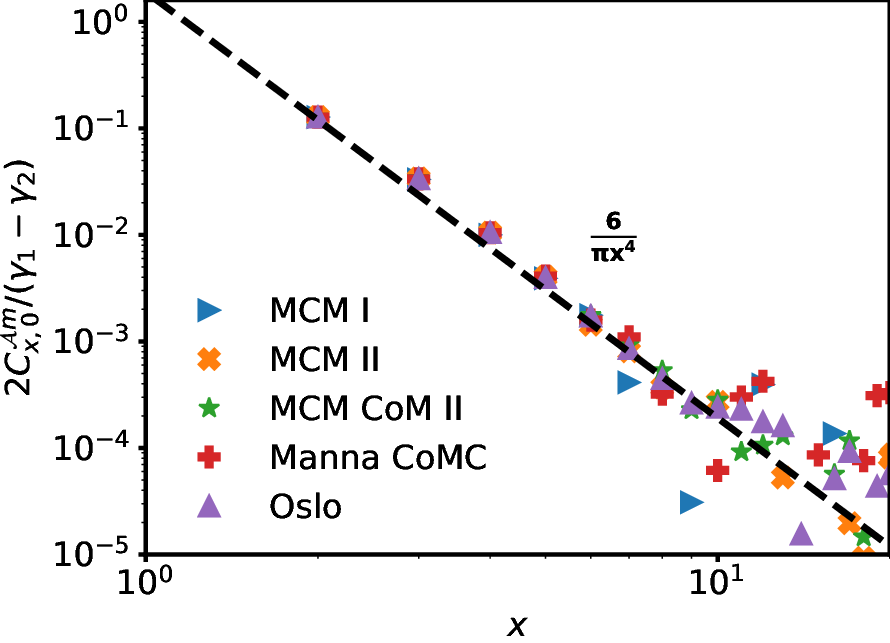}
    \put(-100,150){\textbf{(b)}}
    
    \caption{Two-point correlations in systems with mass conserving, {\it multidirectional} hopping on a two-dimensional periodic square lattice. {\it Panel (a):} Steady-state scaled density-density correlation function, $2DC_{x, 0}^{mm}/(\gamma_1-\gamma_2)$, is plotted as a function of axial distance $x$. 
    {\it Panel (b):} Scaled activity-density correlation function, $2C_{x, 0}^{{\cal A}m}/(\gamma_1 - \gamma_2)$, as a function of spatial distance $x$ is plotted for the Oslo model \cite{Grassberger2016Oct} and Manna model with CoMC \cite{Hexner2017Jan}; here $D = \mathsf{a}'(\rhobar)/2$, $\gamma_1(\rhobar)$ and $\gamma_2(\rhobar)$ all depend on global density $\rhobar$, where $\mathsf{a}(\rhobar)$ is the density-dependent activity $-$ the ``order parameter'' for conserved sandpiles \cite{Dickman1998May}.
     In mass chipping models (MCMs) \cite{Bondyopadhyay2012Jul}, $D$ is independent of density. 
    Theory (black dotted line) predicts a generic power law $(6/\pi) x^{-4}$ [Eq. \eqref{eq:scaled_crur_intro}]. In simulations, we take a periodic $250 \times 250$ square lattice and global density $\rhobar=4.0$ (far above the critical density where absorbing-phase transition occurs for the Manna and Oslo models). 
 }
    \label{fig:corrl}
\end{figure}

{\it Classification of models.$-$} Indeed, based on the calculations of various transport coefficients $\gamma$'s in the previous sections and their relationship to the coefficients $S_0$, $S_1$ and $S_2$ appearing in the structure factor in the small$-{\bf q}$ limit, one can categorize various possible scenarios as given below. To this end, we consider a broad class of models intensively studied in the past several decades: 
\\\\
(i) Symmetric simple exclusion processes (SSEPs) \cite{Derrida2001Sep} $-$ a {\it single} particle hops to a randomly chosen nearest-neighbor site if the site is vacant; 
\\\\
(ii) zero range processes (ZRPs) \cite{Evans2004Jun} $-$ a {\it single} particle hops to a randomly chosen nearest-neighbor site; 
\\\\
(iii) mass chipping models (MCMs) \cite{Aldous1995Jun, Krug2000Apr, Rajesh2000May, Bondyopadhyay2012Jul} $-$ random fractions of (continuous) mass at a site hop to randomly chosen nearest-neighbor sites; 
\\\\
(iv) conserved sandpiles  \cite{Manna1991Apr, Grassberger2016Oct} $-$ two (or more) particles from an {\it active} site hop to randomly chosen nearest-neighbor sites. 
\\
\\
Categories (ii), (iii) and (iv) have {\it no} hardcore constraint. Also, only the categories (iii) and (iv) have {\it multidirectional} hopping of masses, i.e., at each elementary (infinitesimal) time step, mass from the departure site gets redistributed to {\it at least two} different destination sites simultaneously. We consider several variants of MCMs and sandpiles (see Fig. \ref{fig:corrl}), some of which may have an additional {\it center-of-mass conservation} (CoMC) and consequently exhibit hyperuniformity \cite{Torquato2018Jun}. In all cases, we analytically calculate the structure factors, which are expressed in terms of the transport coefficients $\gamma$'s (for details, see Table \ref{tab:Gam_bq_table}). Now, according to Eq. \eqref{eq:Bq_all}, we have the following representative scenarios. 
\begin{figure}
    \centering
    \includegraphics[width=1.0\linewidth]{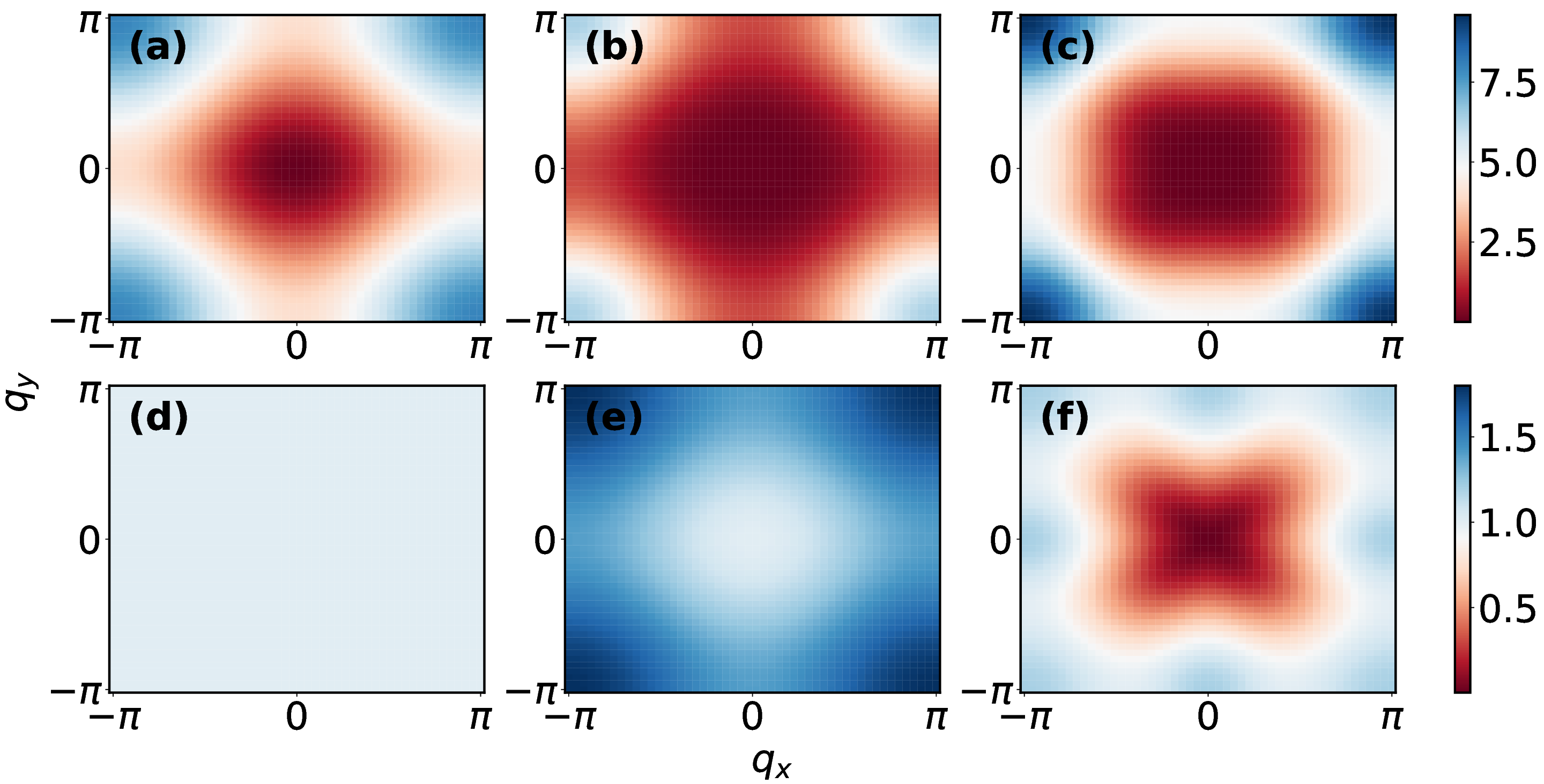}
    \caption{In Panels (a)-(c) and (d)-(f), the quantity $\mathcal{B}(\textbf{q})$ and scaled structure factor $2D S(\textbf{q})$, respectively,  as defined in Eq. \eqref{eq:sq}, are plotted for three representative sets of the Onsager transport coefficients (equivalently, the mobility tensor) $\gamma$'s as defined in Eqs. \eqref{eq:Gqaa} and \eqref{eq:Gqab}. Panels (a) and (d) are for SSEP- or ZRP-like systems (delta-correlated) where $\gamma_0 \neq 0$ and $\gamma_1 = \gamma_2 =0$. Panels (b) and (e) represent the cases where $\gamma_0 = 0$ and $\gamma_1 = \gamma_2\neq 0$; these systems exhibit short-ranged correlations. Panels (c) and (f) illustrate the cases where $\gamma_0 = 0$ and $\gamma_1 > \gamma_2=0$; these systems exhibit power-law correlations $\sim 1/r^{d+2}$. }
    \label{fig:heat_bq}
\end{figure}
\\\\
{\it Case 1:} Unidirectional hopping, $\gamma_0\ne 0$ and $\gamma_1=\gamma_2=0$; this case could correspond to with or without detailed balance and correlations $C^{mm}({\bf r})$ and  $C^{{\cal A}m}({\bf r})$ are short ranged, e.g., SSEP, ZRP, and some variants of MCMs (results not presented).
\\\\
{\it Case 2:} Multidirectional hopping, $\gamma_0\ne 0$ and $\gamma_1 = \gamma_2 \ne 0$; in this case, detailed balance is violated, and correlations $C^{mm}({\bf r})$ and  $C^{{\cal A}m}({\bf r})$ are short-ranged, e.g., some variants of MCMs (results not presented).
\\\\
{\it Case 3:} Multidirectional hopping, $\gamma_0\ne 0$ and $\gamma_2\neq \gamma_1$; detailed balance is violated, power-law correlations with $C^{mm}({\bf r})$ and  $C^{{\cal A}m}({\bf r})$ as $\sim r^{-(d+2)}$, are observed, e.g., MCM I. 
\\\\
{\it Case 4:} Multidirectional hopping, $\gamma_0=0$; the systems are hyperuniform \cite{Torquato2003Oct}, e.g., CoM-conserving models, and correlations $C^{mm}({\bf r})$ and  $C^{{\cal A}m}({\bf r})$ are short ranged  for $\gamma_2 = \gamma_1$ (e.g., MCM CoMC I) and power law otherwise (e.g., MCM-CoMC II, Manna model with CoMC and Oslo model as in Fig. \ref{fig:corrl}).
\\\\
To verify the above theoretical predictions, we have performed Monte-Carlo simulations of various microscopic models in two dimensions.
In panels (a) $-$ (c) and (d) $-$ (f) of Fig. \ref{fig:heat_bq}, we first present the theoretically obtained heat maps of the quantity $2D\mathcal{B}(\textbf{q})$ and the scaled structure factor $2D S(\textbf{q})$, respectively, in the two-dimensional $\textbf{q}$-plane for different sets of $\gamma \equiv \{\gamma_0, \gamma_1, \gamma_2\}$ values.
The non-analyticity emerges when the condition $\gamma_1\neq\gamma_2$ is satisfied. 
We also construct model-independent scaled density-density and activity-density correlation functions $2DC^{mm}_\textbf{r}/(\gamma_1-\gamma_2)$ and $2C^{\mathcal{A}m}_\textbf{r}/(\gamma_1-\gamma_2)$, respectively, which are plotted as a function of spatial distance $\textbf{r}=(x, 0)$ in panels (a) and (b) of Fig. \ref{fig:corrl} and are compared with theory. We take global density $\rhobar=4.0$ [far from (above) the absorbing-phase transition points for the Manna model with CoMC and Oslo model], and a periodic square lattice of area $L \times L = 250\times 250$. The black-dotted line in both panels demonstrates that the scaled correlations indeed exhibit a generic power-law decay $(6/\pi) x^{-4}$ along $x-$axis [as obtained in Eq.~\eqref{eq:scaled_crur}]. Indeed, we see a reasonably good agreement between simulations and theory.

\section{Conclusions}\label{sec:conclusion}

In this paper, we have provided a theoretical characterization of spatial structures in a broad class of conserved-mass transport processes defined on a periodic $d-$dimensional ($d>1$) hypercubic lattice. The models are governed by continuous-time Markov jump processes and, importantly, involve {\it multidirectional} hopping, which respects all symmetries of the lattice. Multidirectional hopping refers to the mass transfer rule where several chunks of mass, or several particles, can hop out simultaneously from a site to the neighboring ones. 
We focus on the often-studied two-point spatial correlation functions involving density and activity in mass chipping models (MCMs) and sandpiles. 
We show that, at large distance $r=|{\bf r}|$, these models exhibit $1/r^{(d+2)}$ power-law correlations in higher dimensions for generic parameter values. In fact, such correlations are observed even when the systems, such as the conserved sandpiles, are at a density much above the critical density where the systems undergo an absorbing-phase transition.

The crucial mechanism behind the algebraic decay is the multidirectional hopping dynamics, where multiple chunks of mass or particles can simultaneously hop out of a lattice site in various directions, thereby breaking detailed balance and generically leading to long-ranged correlations.
By using a hydrodynamic and microscopic approach, the strength of the correlation functions have been determined in terms of the bulk-diffusion coefficient and the Onsager matrix (or, the mobility tensor). More specifically, for the class of models considered here, we parametrize the transport properties, and consequently the correlations, in terms of three coefficients (density-dependent in general), $\gamma_0$, $\gamma_1$ and $\gamma_2$ [see Eq. \eqref{eq:Bq_all}].
Indeed, the presence (or, absence) of the power-law correlations can be understood as follows. 
For $\gamma_1 \ne \gamma_2$, irrespective of the values of $\gamma_0$, the correlation functions decay as a power law [$\sim 1/r^{(d+2)}$]. However, in some special cases, where $\gamma_1=\gamma_2$, and therefore $S_2=0$ [see Eq.~\eqref{Sq} and \eqref{eq:s2}], the correlation functions can be short-ranged too; for a schematic ``phase diagram'', see Fig. \ref{fig:phase_diagrame}. 
\begin{figure}
\centering
\begin{tikzpicture}[scale=1]
\draw[->] (0,0) -- (5.8,0);
\node at (5.5, -0.2)  {$\gamma_1$};
\draw[->] (0,0) -- (0,5.8) node[anchor=south east] {$\gamma_2$};



\fill[blue!50, opacity=0.5] (0,0) -- (5.5,5.5) -- (0,5.5) -- cycle; 
\fill[red!50, opacity=0.5] (0,0) -- (5.5,5.5) -- (5.5,0) -- cycle;  

\draw[dashed] (0,0) -- (5.5,5.5);
\node[rotate=45] at (2.5,2.8) {$\textcolor{red}{\gamma_1 = \gamma_2}$(Short-ranged)};
\node[fill=white] at (2.0, 4.0) {\textbf{Power law}};
\node[fill=white] at (3.5, 1.0) {\textbf{Power law}};

\end{tikzpicture}
\caption{Schematic ``phase diagram'' of conserved-mass transport processes on a $d-$dimensional lattice in the plane of parameters $\gamma_1$ and $\gamma_2$ as defined in Eqs. \eqref{eq:Gqaa} and \eqref{eq:Gqab}.
The black-dotted line depicts the ``phase boundary'', where the correlation function is short-ranged, and elsewhere it is a power-law.  }
\label{fig:phase_diagrame}
\end{figure}

In particular, our theory explains why center-of-mass-conserving dynamics with axial bidirectional (opposite) hopping of masses generically leads to long-ranged correlations, which manifest into a remarkable state of nonequilibrium disordered matter, called {\it hyperuniformity} \cite{Torquato2003Oct, Hexner2017Jan}. In that case, we have $\gamma_0=\gamma_2=0$, but the coefficient of the {\it nonanalytic} term in static structure factor is still nonzero as $\gamma_1 \ne 0$ [implying $S_2 \ne 0$; see Eqs. \eqref{Sq} and \eqref{eq:s2}]. Therefore, the systems have anomalously suppressed long-wave-length density fluctuations $S({\bf q} \to 0)=0$, resulting in hyperuniformity [$S(q) \sim q^{\alpha}$ with hyperuniformity exponent $\alpha=2$] and long-ranged correlation functions [$\sim 1/r^{(d+2)}$] at the same time.

Importantly, the systems studied in this work are isotropic (i.e., symmetric under lattice rotations) and homogeneous, yet they violate detailed balance. Moreover, the symmetric nature of mass hopping ensures that there is no net current in real space (though probability current in the configuration space is nonzero), and the systems exhibit diffusive relaxation at large spatio-temporal scales. 
While such systems were widely studied over the past several decades \cite{Aldous1995Jun, Dickman2001Oct, Bondyopadhyay2012Jul, Grassberger2016Oct}, so far it has not been realized that they possess generic scale invariance in higher dimensions, even far from the (absorbing-)phase transition point, if any.

In contrast,  in open current-carrying systems, e.g., in the presence of a temperature or density gradient (therefore, anisotropic) \cite{Procaccia1979Jan,  Kirkpatrick1979Apr, Law1988Apr, Derrida2005Mar}, long-ranged correlations are rather the rule \cite{Spohn1983Dec, Zhang1988Sep, Hwa1989Apr, Doyon2023Jul, Bertini2007Jul} (in equilibrium, apart from the critical state, continuous-symmetry-broken ordered phase can also have long-ranged correlations \cite{Wang2024Sep}). Furthermore, it is known that closed systems with localized drive (``disorder'') can induce long-ranged correlations \cite{Sadhu2014Jul}. But, this raises an important question: What happens in closed, homogeneous and isotropic systems? Is the violation of detailed balance sufficient to generate power laws, even on a lattice that has more restricted symmetry compared to that in a continuum? 
The answers to these questions are not quite obvious.
Though it was previously explained why two-point correlations in closed {\it anisotropic} systems have $1/r^d$ decay \cite{Maes1990Nov, Grinstein1990Apr, Garrido1990Aug, vanBeijeren1990Sep, Maes1991Sep, Dorfman1994Oct},
the behavior of the correlations in systems when all symmetries of the lattice are present is not well understood. 
Indeed, one prior attempt to address this issue used an approximate kinetic theory to demonstrate that an isotropic lattice model too can exhibit algebraic decay, albeit a faster one $1/r^{d+2}$ on a hypercubic lattice \cite{Bussemaker1996Jun}. 
However, a detailed characterization of the class of microscopic dynamics for which such decays occur is still lacking.
In this scenario, here we elucidate, in terms of the Onsager transport coefficients, its precise dynamical mechanism, that crucially requires {\it multidirectional} hopping to violate detailed balance and generate algebraic decay.
From a general perspective, our findings could shed light on why power laws are so ubiquitous in nature \cite{Bak1987Jul}
$-$ the question posed by Bak, Tang, and Wiesenfeld in their seminal paper more than three decades ago.

\section{Acknowledgment} 

We thank Deepak Dhar and Christian Maes for their useful discussion and insightful comments on the manuscript. We also thank Dhiraj Tapadar for discussion.

\appendix
\widetext
\section{Fluctuating current correlations}

The time-integrated bond current up to time $t$ along a given direction $\alpha$ can be written as 
\begin{align}
    \cQ_\alpha(\textbf{r}, t) = \int_{0}^{t}dt' j_\alpha(\textbf{r},t'),
\end{align}
where $j_\alpha(\textbf{r},t)$ is
instantaneous current, which can be written as a sum of diffusive and fluctuating current components,
\begin{align}
    j_\alpha(\textbf{r},t) = j_\alpha^{(d)}(\textbf{r},t)+ j_\alpha^{(fl)}(\textbf{r},t).
\end{align}
Now, using the infinitesimal update rules given in the main text, the unequal-time bond current correlation in directions $\alpha$ and $\beta$ can be written as
\begin{align}\label{eq:cqcqabttp_ap}
    \frac{d}{dt}C_{\textbf{r}}^{\cQ_\alpha\cQ_\beta}(t,t') = C_{\textbf{r}}^{j^{(d)}_\alpha\cQ_\beta}(t,t').
\end{align}
To solve the above equation, we require equal-time current-current correlation, which satisfies the following equation,
\begin{align}\label{eq:cqcqabtt_ap}
    \frac{d}{dt}C_\textbf{r}^{\cQ_\alpha\cQ_\beta}(t, t) = C_\textbf{r}^{j^{(d)}_\alpha\cQ_\beta}(t, t)+C_\textbf{r}^{\cQ_\alpha j^{(d)}_\beta}(t, t)+\Gamma^{\alpha\beta}_{\textbf{r}}.
\end{align}
The first two terms of the right hand of the above equation correspond to an infinitesimal change of bond current either in $\alpha$ or in $\beta$ directions, whereas the third term $\Gamma^{\alpha\beta}_{\textbf{r}}$ represents the simultaneous update of current in both directions ($\alpha$ and $\beta$). Using Eqs. \eqref{eq:cqcqabtt_ap}  and \eqref{eq:cqcqabttp_ap}, we obtain the solution of unequal-time current correlations as given below:
\begin{align}
    C_{\textbf{r}}^{\cQ_\alpha\cQ_\beta}(t,t') =  \Gamma_{\textbf{r}}^{\alpha\beta}t' + \int_0^{t'}dt''\big[C_\textbf{r}^{j^{(d)}_\alpha\cQ_\beta}(t'', t'')+C_\textbf{r}^{\cQ_\alpha j^{(d)}_\beta}(t'', t'')\big]+\int_{t'}^{t}dt'' C_{\textbf{r}}^{j^{(d)}_\alpha\cQ_\beta}(t'',t').
\end{align}
Now, we can write the time-integrated bond-current correlation fluctuation in the following form, 
\begin{align}\label{eq:crqaqbttp}
    C_{\textbf{r}}^{\cQ_\alpha\cQ_\beta}(t,t') = \Theta(t-t')C_{\textbf{r}}^{\cQ_\alpha\cQ_\beta}(t,t')
    &+\Theta(t'-t)C_{\textbf{r}}^{\cQ_\alpha\cQ_\beta}(t',t),
\end{align}
where the Heaviside step function $\Theta (t) \text{ equals } 1, 1/2 \text{ and } 0$ for $t>0$, $t=0$ and $t<0$, respectively.
First differentiating Eq. \eqref{eq:crqaqbttp} w.r.t. $t'$ and then by $t$ we have
\begin{align}\label{eq:cjajbttp}
    C_\textbf{r}^{j_\alpha j_\beta}(t,t')= \frac{d}{dt}\frac{d}{dt'} C_{\textbf{r}}^{\cQ_\alpha\cQ_\beta}(t,t') = \Gamma^{\alpha\beta}_\textbf{r}\delta(t-t')+\Theta(t-t')C_{\textbf{r}}^{j^{(d)}_\alpha j_\beta}(t, t') +\Theta(t'-t)C_{-\textbf{r}}^{j^{(d)}_\beta j_\alpha}(t', t) .
\end{align}
Furthermore, the unequal-time fluctuating current correlation can be written as
\begin{align}\label{eq:crflttp}
    C_\textbf{r}^{j^{(fl)}_\alpha j^{(fl)}_\beta}(t,t') 
    =  C_\textbf{r}^{j_\alpha j_\beta}(t, t') -  C_\textbf{r}^{j^{(d)}_\alpha j_\beta}(t, t'),
\end{align}
where we have used $\langle j_\alpha ^{(fl)}j^{(d)}_\beta(t,t')\rangle_c = 0$ for $t>t'$,  implying that the diffusive current at the initial time $t'$ is uncorrelated with the fluctuating current at the latter time $t$.  Finally, using Eq. \eqref{eq:cjajbttp} in Eq. \eqref{eq:crflttp},  we obtain the fluctuating current correlation function,
\begin{align}
    C_\textbf{r}^{j^{(fl)}_\alpha j^{(fl)}_\beta}(t,t') = \Gamma^{\alpha\beta}(\textbf{r})\delta(t-t'),
\end{align}
which is Eq. \eqref{eq:jflabttp} in the main text.

\twocolumngrid
\bibliographystyle{apsrev4-2}
\bibliography{references}


\end{document}